\documentclass[a4paper,fleqn]{cas-sc}

\usepackage[authoryear,longnamesfirst]{natbib}
\usepackage{float}
\usepackage{etoolbox}
\AtBeginEnvironment{highlights}{\comment}
\AtEndEnvironment{highlights}{\endcomment}

\restylefloat{figure}
\usepackage{graphicx}

\usepackage{xcolor}

\def\tsc#1{\csdef{#1}{\textsc{\lowercase{#1}}\xspace}}
\tsc{WGM}
\tsc{QE}

\begin{document}

\shorttitle{}    

\shortauthors{Alhassan et al.}  

\title [mode = title]{COVID-19 Forecasting from U.S. Wastewater Surveillance Data: A Retrospective Multi-Model Study (2022--2024)}  

\tnotemark[1] 


\author[1,2]{Faharudeen Alhassan}
\fnmark[1]
\cormark[1]
\ead{falhassan@gsu.edu}



\affiliation[1]{organization={Department of Mathematics and Statistics, Georgia State University},
            city={Atlanta},
            state={GA},
            postcode={30303}, 
            country={USA}}

\author[1,2]{Hamed Karami}
\ead{hkarami@gsu.edu}

\credit{}

\author[2]{Amanda Bleichrodt}
\ead{ableichrodt@gsu.edu}
\credit{}

\author[5]{James M. Hyman}
\ead{mhyman@tulane.edu}
\credit{}

\author[3]{Isaac C. H. Fung}
\ead{cfung@georgiasouthern.edu}

\author[2]{Ruiyan Luo}
\ead{rluo@gsu.edu}
\credit{}

\author[2,4]{Gerardo Chowell}
\cormark[1]
\ead{gchowell@gsu.edu}
\credit{}

\affiliation[3]{
  organization = {Department of Biostatistics, Epidemiology and Environmental Health Sciences, Jiann-Ping Hsu College of Public Health, Georgia Southern University},
  city={Statesboro},
  state={GA},
  postcode={30460},
  country={USA}
}

\affiliation[5]{
  organization = {Department of Mathematics, Tulane University},
  city={New Orleans},
  state={LA},
  postcode={70118},
  country={USA}
}
\affiliation[2]{organization={Department of Population Health Sciences, Georgia State University},
            city={Atlanta},
            state={GA},
            postcode={30303}, 
            country={USA}}

\affiliation[4]{organization={Department of Applied Mathematics, Kyung Hee University},
            city={Yongin},
            state={GA},
            postcode={17104}, 
            country={Korea}}

\cortext[1]{Corresponding author: gchowell@gsu.edu}

\begin{abstract}
Accurate and reliable forecasting models are critical for guiding public health responses and policy decisions during pandemics such as COVID-19. Retrospective evaluation of forecasting performance provides an essential framework for assessing and improving epidemic prediction methods. In this study, we used COVID-19 wastewater data from CDC's National Wastewater Surveillance System to generate sequential weekly retrospective out-of-sample forecasts for the United States from March 2022 through September 2024, both at the national level and for four major regions (Northeast, Midwest, South, and West). We produced 133 weekly forecasts using 11 models, including ARIMA, generalized additive models (GAM), simple linear regression (SLR), Prophet, and the \(n\)-sub-epidemic framework (top-ranked, weighted-ensemble, and unweighted-ensemble variants). Forecast performance was assessed using mean absolute error (MAE), mean squared error (MSE), weighted interval score (WIS), and 95\% prediction interval coverage. The \(n\)-sub-epidemic unweighted ensembles outperformed all other models at 3--4-week horizons, particularly at the national level and in the Midwest and West. ARIMA and GAM performed best at 1--2-week horizons in most regions, whereas Prophet and SLR consistently underperformed across regions and horizons. These findings highlight the value of region-specific modeling strategies and demonstrate the utility of the \(n\)-sub-epidemic framework for real-time outbreak forecasting using wastewater surveillance data.
\end{abstract}

\begin{highlights}
\item Retrospective COVID-19 forecasts were generated weekly using CDC NWSS wastewater data at national and regional scales.  
\item Eleven models (ARIMA, GAM, SLR, Prophet, \(n\)-sub-epidemic variants) were compared across 1--4-week horizons.  
\item The unweighted \(n\)-sub-epidemic ensemble delivered the most accurate long-term forecasts, especially in the Midwest and the West.  
\item ARIMA and GAM excelled at short-term forecasts in South and Northeast; Prophet and SLR consistently underperformed.  
\item Region-specific modeling enhances forecasting reliability and informs tailored public health responses.  
\end{highlights}

\begin{keywords}
COVID-19 \sep Wastewater surveillance \sep Ensemble modeling \sep \(n\)-sub-epidemic framework \sep Time series forecasting
\end{keywords}

\maketitle

\section{Introduction}
The COVID-19 pandemic has highlighted the urgent need for efficient, scalable, and timely disease surveillance methods. Traditional epidemiological approaches rely primarily on clinical testing and hospital records, which, while valuable, are often constrained by delays in reporting, limited testing availability, and the under-detection of asymptomatic infections \citep{nixon2022real,alvarez2023limitations}. These limitations have underscored the importance of alternative surveillance methods, especially wastewater-based epidemiology (WBE), which has emerged as a cost-effective, non-invasive, and real-time tool to monitor SARS-CoV-2 transmission at the community level \citep{medema2020presence,kitajima2020sars,parkins2024wastewater,rashid2024scoping}. Viral particles are shed in human waste even before individuals exhibit symptoms or seek medical attention \citep{medema2020presence,kitajima2020sars}. Unlike traditional testing, which depends on individual participation and healthcare access, WBE provides a comprehensive snapshot of community-wide infection levels. As a result, it has been implemented in a wide range of settings internationally to track SARS-CoV-2 dynamics \citep{rashid2024scoping,parkins2024wastewater}. Because wastewater signals often precede clinical indicators, WBE has increasingly been explored as a complementary data stream to support epidemic monitoring and short-term forecasting efforts.

Recent research has demonstrated that SARS-CoV-2 RNA concentrations in wastewater are often associated with reported COVID-19 cases, hospitalizations, and other epidemiologic indicators \citep{peccia2020measurement,weidhaas2021correlation,li2023correlation}. This relationship establishes WBE as an effective early warning system, allowing public health officials to detect emerging outbreaks and monitor infection trends \citep{gonzalez2020covid}. At the same time, the strength and timing of these relationships are not uniform across locations or time periods. Systematic review evidence has shown substantial variability in wastewater--clinical correlations across studies, with estimates influenced by catchment characteristics, sampling frequency, environmental conditions, and clinical testing coverage \citep{li2023correlation}. In addition, these relationships may vary across dominant variant periods and changes in healthcare-seeking or testing behavior \citep{zhan2023correlative}. WBE has been successfully applied to epidemic forecasting on different geographical scales, from localized wastewater treatment plants to national surveillance networks \citep{joseph2022assessing,li2023wastewater}. For instance, a study in Catalonia (Spain) demonstrated the potential of WBE to predict SARS-CoV-2 incidence across multiple regions, reinforcing its role in epidemiological forecasting \citep{joseph2022assessing}. Similarly, research in the United States (U.S.) showed that wastewater data could predict weekly COVID-19 hospital admissions up to four weeks in advance, emphasizing its value in public health preparedness \citep{li2023wastewater}. In rural communities where clinical testing is often limited, WBE has proven particularly beneficial. In a recent multi-state analysis of 189 wastewater treatment plants during 2023--2024, it was demonstrated that SARS-CoV-2 RNA concentrations precede hospitalization admissions by 2 to 12 days, with lead time varying by state-level WBE coverage and facility characteristics \citep{schenk2024sars}. A study in Idaho successfully integrated wastewater data with susceptible--exposed--infectious--recovered (SEIR) modeling to forecast outbreaks with lead times of up to 11 days, showcasing its ability to enhance surveillance in underserved areas \citep{meadows2025epidemiological}. However, most of these studies have focused on short-term forecasts or single geographic areas, leaving a gap in understanding the comparative performance of forecasting models applied retrospectively across diverse U.S. regions. With the gradual de-escalation of widespread clinical testing and case-based reporting in the U.S., wastewater epidemiology is becoming an increasingly important data source for forecasting potential public health impacts \citep{nixon2022real,parkins2024wastewater}. This shift further underscores the need to evaluate the forecasting performance of models applied to wastewater data across different geographic and temporal contexts.

Retrospective forecasting plays a crucial role in evaluating the performance of COVID-19 prediction models across diverse settings. By systematically comparing past forecasts with actual observed outcomes, researchers can refine methodologies, optimize model parameters, and improve predictive accuracy \citep{drews2022model}. Such evaluations are essential for identifying sources of uncertainty, enhancing forecast calibration, and determining the effectiveness of various statistical and mechanistic modeling approaches \citep{colonna2022retrospective}. Despite these advancements, many forecasting models struggle to account for abrupt shifts in outbreak dynamics, such as sudden spikes in infections or unexpected declines in hospitalizations, critical changes that directly influence public health decision-making \citep{lopez2024challenges}. Improving the ability of models to detect these turning points is crucial for enhancing forecasting systems and ensuring timely interventions. Integrating dynamic modeling approaches with wastewater surveillance data has been proposed as a promising strategy to enhance predictive accuracy and real-time outbreak monitoring \citep{phan2023making,proverbio2022model}.

In this study, we conducted a comprehensive retrospective out-of-sample evaluation of 1- to 4-week forecast performance across 11 different models. Our analysis includes \textit{n}-sub-epidemic models alongside statistical approaches such as autoregressive integrated moving average (ARIMA), generalized additive models (GAMs), generalized logistic growth models (GLMs), simple linear regression (SLR), and Meta (Facebook's) Prophet model. We assess forecasting performance at both national and regional levels, covering four U.S. regions: the Midwest, Northeast, South, and West. The evaluation period spans from March 5, 2022, to September 14, 2024. To ensure a rigorous and standardized assessment, we utilize key epidemic forecasting performance metrics, including mean absolute error (MAE), mean squared error (MSE), 95\% prediction interval (PI) coverage, and weighted interval score (WIS). By systematically comparing multiple forecasting approaches using a rolling-origin retrospective framework, our study provides insights into the relative strengths of statistical and phenomenological models for short-term epidemic forecasting using wastewater surveillance data.

\section{Methods\label{sec-model}}

\subsection{Data Source and Preparation \label{sec}}
The wastewater data used for forecasting COVID-19 trends was obtained from the National Wastewater Surveillance System (NWSS), which is hosted by the Centers for Disease Control and Prevention (CDC)~\citep{CDC}. Wastewater data provides an early warning system, indicating whether infection levels in a particular region may be increasing or decreasing.

The dataset covers national-level data and data from four regions: Midwest, Northeast, South, and West. The regions are defined as follows:

\begin{itemize}
    \item \textbf{West}: Alaska, Arizona, California, Colorado, Guam, Hawaii, Idaho, Montana, Nevada, New Mexico, Oregon, Utah, Washington, Wyoming (N=14).
    \item \textbf{Midwest}: Illinois, Indiana, Iowa, Kansas, Michigan, Minnesota, Missouri, Nebraska, North Dakota, Ohio, South Dakota, Wisconsin (N=12).
    \item \textbf{Northeast}: Connecticut, Maine, Massachusetts, New Hampshire, New Jersey, New York, Pennsylvania, Puerto Rico, Rhode Island, Vermont (N=10).
    \item \textbf{South}: Arkansas, Alabama, Delaware, District of Columbia, Florida, Georgia, Kentucky, Louisiana, Maryland, Mississippi, North Carolina, Oklahoma, South Carolina, Tennessee, Texas, Virginia, and West Virginia (N=17).
\end{itemize}

Wastewater viral activity levels are used to assess the concentration of the virus in wastewater, which may indicate the infection risk in a given area. These levels are categorized as follows:

\begin{itemize}
    \item \textbf{Minimal}: Up to 1.5
    \item \textbf{Low}: 1.5--3
    \item \textbf{Moderate}: 3--4.5
    \item \textbf{High}: 4.5--8
    \item \textbf{Very High}: Greater than 8
\end{itemize}

The viral activity level is calculated based on the number of standard deviations above the baseline and transformed to a linear scale using the formula:
\[
\text{Wastewater Viral Activity Level (WVAL)} = e^{x},
\]
 where $x$ is the number of standard deviations relative to baseline \citep{CDC1}.
 
It is important to note that WVAL represents a normalized metric rather than absolute viral concentration, allowing for comparison across different wastewater treatment facilities with varying population sizes and sampling protocols.

Wastewater samples were collected by state, tribal, local, and territorial partners as part of CDC's National Wastewater Surveillance System (NWSS), following CDC guidance on site selection, sampling, and laboratory methods \citep{kirby2021using}. The network expanded from 209 sampling sites in September 2020 to more than 1{,}500 by December 2022, covering roughly 47\% of the U.S. population \citep{adams2024national}. Catchment sizes range from systems serving thousands of residents to large utilities serving over one million people, and participating laboratories implemented standardized methods and quality controls recommended by CDC and NWSS partners \citep{adams2024national, kirby2021using}

\begin{figure}[htbp]
\centering
\includegraphics[width=0.8\textwidth]{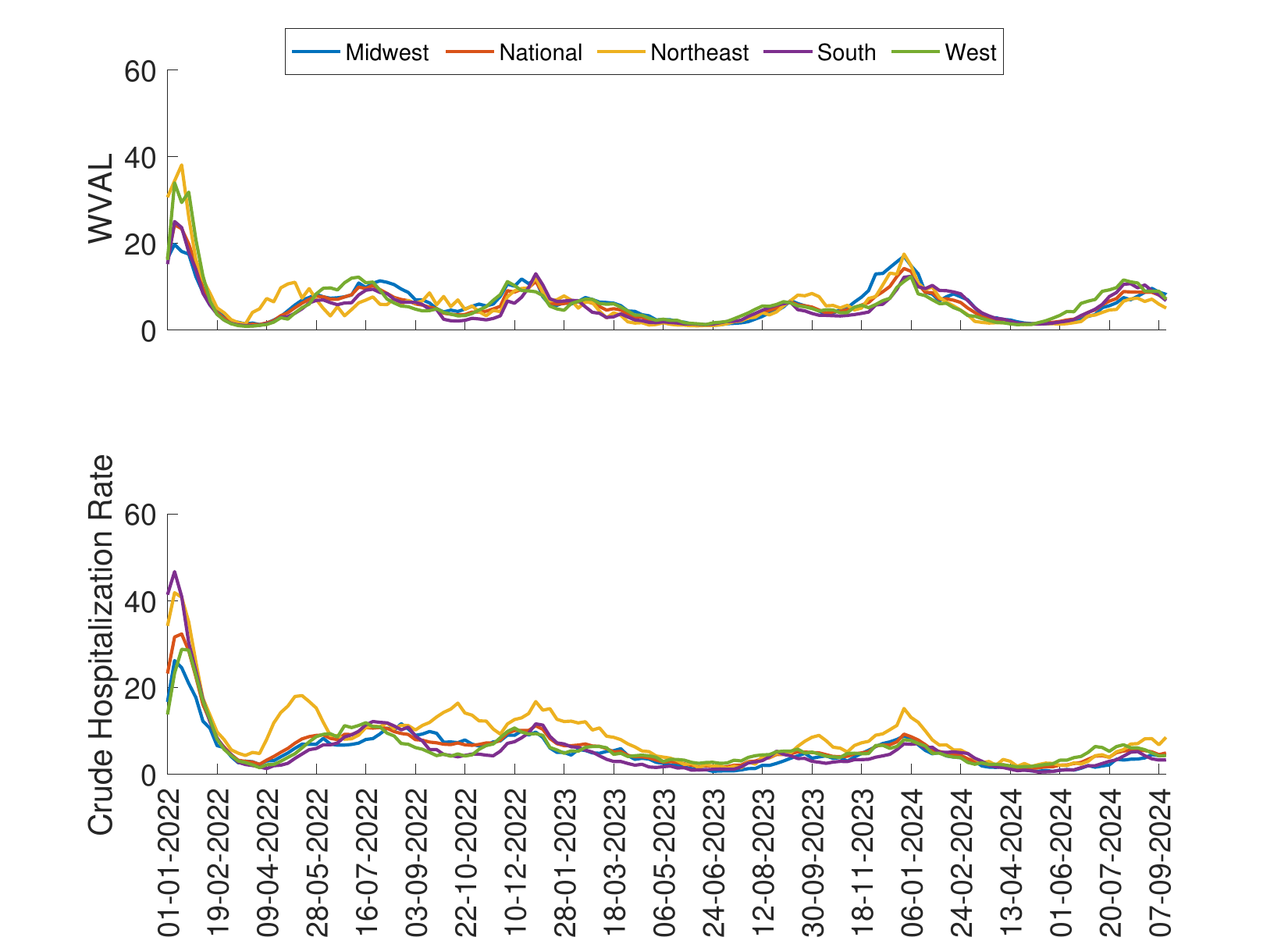}
\caption{\footnotesize{Time-series comparison of wastewater viral activity level (WVAL) and crude COVID-19 hospitalization rates from January 2022 to September 2024. The top panel illustrates the WVAL detected in wastewater surveillance data, and the bottom panel represents the crude hospitalization rate over the same period. Both datasets exhibit periodic fluctuations, with notable peaks aligning at several time points, suggesting a potential correlation between wastewater viral activity and hospital admissions. The observed trends indicate that increases in wastewater viral concentration often precede corresponding surges in hospitalization rates, highlighting the potential utility of wastewater surveillance as an early warning system for monitoring COVID-19 outbreaks. This relationship emphasizes the importance of integrating wastewater-based epidemiology into public health decision-making. The lead time between WVAL peaks and hospitalization surges appears to vary by region and outbreak intensity, typically ranging from 2-12 days based on recent U.S. 
surveillance data \citep{schenk2024sars}
}}
\label{fig:1}
\end{figure}

\subsection{Model calibration and specification}

We used a total of 11 models in our analysis, including the top-ranked models and ensemble models (weighted and unweighted) from the \textit{n}-sub-epidemic framework, together with autoregressive integrated moving average (ARIMA), generalized additive models (GAMs), generalized logistic growth models (GLMs), simple linear regression (SLR), and Prophet models. A detailed description of these models is provided below.

We analyzed the CDC National Wastewater Surveillance System (NWSS) weekly wastewater viral activity level (WVAL) time series available in the October 26, 2024 CDC release at the national level and for four U.S. regions. In this study, ``all data'' refers to the complete sequence of weekly published WVAL values from January 1, 2022, through September 14, 2024, which was used only to define the retrospective sequence of forecast origins and evaluation targets. We used the published weekly national and regional WVAL series as provided in that CDC/NWSS release and did not reconstruct these series from raw site-level wastewater measurements. No additional rolling average, interpolation, or site-level reweighting was applied.

For each forecast origin, models were fit only to the most recent 10 weeks of data available at that time. Starting on March 5, 2022 (10 weeks after the first observation in the study period), we advanced the forecast origin by one week at a time, re-estimated all models using the rolling calibration window $[t-9,t]$, and generated 1- to 4-week-ahead forecasts for weeks $t+1$ through $t+4$. Forecast accuracy was then evaluated against the subsequently observed WVAL values. Thus, the full January 2022--September 2024 record was used only to define the sequence of retrospective forecast origins and out-of-sample evaluation targets; no future observations were used in model fitting at any forecast date, and training and evaluation data were kept separate throughout.

Because our analysis was conducted on the published weekly WVAL series rather than raw site-level wastewater measurements, differences in sampling frequency across contributing sites or jurisdictions were not modeled explicitly as separate weights in our forecasting framework. Accordingly, our comparison evaluates model performance on the published weekly surveillance signal rather than on a site-level harmonized dataset. This is an important scope condition and should be considered when interpreting differences in forecast performance across regions and time periods.

The 10-week calibration window was selected to balance having sufficient observations to capture recent local dynamics while limiting the influence of older observations, changing variability, and possible structural breaks in the wastewater signal, consistent with prior short-term epidemic forecasting studies \citep{bleichrodt2024evaluating,bleichrodt2023real}.

All comparison models used only historical WVAL values within the calibration window and did not incorporate external covariates. The \textit{n}-sub-epidemic framework assumes that the observed trajectory can be represented as a superposition of overlapping sub-epidemic waves. ARIMA assumes that short-term dependence in the differenced series can be captured through autoregressive and moving-average components. GAM assume a smooth nonlinear relationship between WVAL and time over the calibration window. GLM assume generalized logistic growth dynamics over the fitting period. SLR assumes an approximately linear trend in WVAL over the calibration window. Prophet assumes that the time series can be represented through decomposable trend components estimated from recent observations.

We assumed a normal error structure in the observed data for the models estimated under this framework. This decision was based on two considerations: (i) the short 10-week calibration window helped reduce the influence of time-varying variance and structural changes, and (ii) the models demonstrated high empirical calibration coverage, suggesting that the normal approximation was adequate over the fitting period.

\subsection{The \textit{n}-sub-epidemic modeling framework}
This framework is a flexible approach used to capture complex epidemic patterns by combining multiple overlapping sub-epidemics \citep{chowell2024subepipredict,chowell2022ensemble}. Rather than viewing an outbreak as a single continuous event, this framework breaks it into smaller, more manageable components, each representing a distinct sub-epidemic. These sub-epidemics may correspond to different geographic regions, demographic groups, or periods within the course of an epidemic.

Each sub-epidemic is modeled using a generalized logistic growth function \citep{chowell2024subepipredict} that captures its growth rate, final size, and scaling behavior. This allows the model to represent a wide range of early outbreak dynamics, from slow, sub-exponential growth to rapid, exponential expansion. The mathematical structure is consistent across all sub-epidemics, but each sub-epidemic is characterized by a unique set of parameters.

An $n$-sub-epidemic trajectory consists of $n$ overlapping sub-epidemics and is described by the system of differential equations
\begin{equation}
\frac{dC_i(t)}{dt} = A_i(t)\, r_i C_i^{p_i}(t)\left(1 - \frac{C_i(t)}{K_{0i}}\right), 
\quad i = 1,\dots,n.
\label{eq:sub}
\end{equation}

In the original $n$-sub-epidemic formulation, $C_i(t)$ denotes the cumulative number of cases associated with the $i$-th sub-epidemic. 
In our wastewater setting, we instead treat $C_i(t)$ as the wastewater viral activity level (WVAL) contributed by the $i$-th sub-epidemic, and 
we use the same dynamical form as a flexible phenomenological model for the rise and fall of the WVAL signal. 
The term $A_i(t)$ is a binary activation indicator (0 or 1) that determines when the $i$-th sub-epidemic is active. 
Under this interpretation, the $i$-th sub-epidemic evolves according to Eq.~\eqref{eq:sub}.

where:
\begin{itemize}
    \item $C_i(t)$ is the wastewater viral activity level (WVAL) contributed by the $i$-th sub-epidemic at time $t$,
    \item $\dfrac{dC_i(t)}{dt}$ is the rate of change in the modeled WVAL signal for the $i$-th sub-epidemic at time $t$,
    \item $r_i C_i^{p_i}(t)$ is the growth term governing how rapidly the wastewater signal increases, with $p_i$ modulating the growth regime,
    \item $\left( 1 - \dfrac{C_i(t)}{K_{0i}} \right)$ is a saturation term that slows growth as $C_i(t)$ approaches the asymptotic level $K_{0i}$.
\end{itemize}

The model parameters have the following interpretations:
\begin{itemize}
    \item $r_i$: intrinsic growth rate of the WVAL signal for the $i$-th sub-epidemic,
    \item $p_i \in [0,1]$: scaling parameter controlling the effective growth regime (with $p_i = 0$ corresponding to approximately constant growth, $p_i = 1$ to exponential growth, and $0 < p_i < 1$ to sub-exponential growth),
    \item $K_{0i}$: asymptotic (plateau) level of the WVAL signal for the $i$-th sub-epidemic.
\end{itemize}

To determine when a new sub-epidemic begins, the model employs an indicator $A_i(t)$ that activates the $(i+1)$-th sub-epidemic once the modeled wastewater viral activity level $C_i(t)$ for the $i$-th sub-epidemic exceeds a specified threshold $C_{\text{thr}}$.
\begin{equation}
A_i(t) =
\begin{cases}
1, & \text{if } C_{i-1}(t) > C_{\text{thr}}, \\
0, & \text{otherwise},
\end{cases} \quad \text{for } i = 2, \ldots, n,
\end{equation}
with $A_1(t) = 1$ to activate the first sub-epidemic by default.

The initial value of observed viral loads is given by:
\begin{equation}
C_1(0) = I_0,
\end{equation}
where $I_0$ is the initial viral load at the start of the outbreak.

The cumulative epidemic curve is defined as the sum of all individual sub-epidemic trajectories:
\begin{equation}
C_{\text{tot}}(t) = \sum_{i=1}^{n} C_i(t).
\end{equation}

When modeling a single sub-epidemic ($n = 1$), the model reduces to the classic three-parameter logistic growth model. For more complex epidemic dynamics (such as multiple peaks or plateaus, as seen in Figure \ref{fig:1}), using multiple sub-epidemics ($n > 1$) increases modeling flexibility and accuracy. A full \textit{n}-sub-epidemic model requires estimation of $3n + 1$ parameters. In this paper, we set \textit{n} $\leq 2$ in the \textit{n}-sub-epidemic trajectory.

 To enhance the accuracy of the estimation and evaluation process, we used 30 initial guesses for the parameter estimation process (\texttt{numstartpoints = 30}). We conducted 300 bootstrap realizations (\texttt{B}) to characterize parameter uncertainty effectively.

The \textit{n}-sub-epidemic modeling framework offers a powerful and adaptable tool for analyzing and forecasting disease outbreaks. It is particularly well-suited to epidemics that display multi-wave progression. This framework has demonstrated excellent performance in modeling various infectious diseases, including Ebola, Zika, COVID-19, and mpox \citep{chowell2024subepipredict,chowell2022ensemble,bleichrodt2024evaluating}.

\subsubsection{Parameter estimation and model selection}
To estimate the model parameters, we used a nonlinear least-squares fitting method, aligning the model's predictions with the observed COVID-19 wastewater data \citep{chowell2024subepipredict}. After fitting several models, each using a different threshold value \( C_{\text{thr}} \) to define when new sub-epidemics begin. The  \( C_{\text{thr}} \) values are generated by uniformly partitioning the cumulative sum of the smoothed signal into several levels equal to the total data points. We evaluated their performance using the corrected Akaike Information Criterion (\( AIC_c \)). This criterion helps identify models that best balance goodness-of-fit and model complexity.

The formula for \( AIC_c \) is:
\begin{equation}
AIC_c = n_d \log(\text{SSE}) + 2m + \frac{2m(m+1)}{n_d - m - 1}, 
\end{equation}
where each component serves a specific purpose in model selection:

\begin{itemize}
    \item $n_d \log(\text{SSE})$: Measures the goodness-of-fit, where SSE is the sum of squared errors between observed and predicted values
    \item $2m$: Penalty term for model complexity to prevent overfitting, where $m$ is the number of parameters
    \item $\frac{2m(m+1)}{n_d - m - 1}$: Finite sample correction that becomes important when the sample size $n_d$ is small relative to the number of parameters
\end{itemize}

The parameters are defined as follows:
\begin{itemize}
    \item \( \text{SSE} \) is the sum of squared errors between the observed data and model predictions,
    \item \( m \) is the number of model parameters,
    \item \( n_d \) is the number of data points.
\end{itemize}

This method assumes that the residuals (errors) in the data follow a normal distribution. More details about the parameter estimation procedure are provided in \citep{chowell2022ensemble}.

\subsubsection{Parametric bootstrapping}
To assess uncertainty in the parameters of the best-fitting model \(f(t,\widehat\Theta)\), we used a parametric bootstrapping approach \citep{hastie2009elements}.  Here \(f(t,\widehat\Theta)\) denotes the model's predicted value at time \(t\) using the estimated parameters \(\widehat\Theta\).  This method involves generating multiple datasets by resampling from the fitted model and then reestimating the parameters for each sample.  This helps quantify variability in parameter estimates and model forecasts without relying on closed-form solutions.

In addition to estimating standard errors and confidence intervals, we used bootstrapping to produce one- to four-week-ahead forecasts with associated uncertainty bounds. In this analysis, we generated 300 bootstrap realizations to characterize the parameter uncertainty and uncertainty in forecasting.

\subsubsection{Building ensemble \( n \)-sub-epidemic models}
The ensemble models were constructed by combining several of the top-performing sub-epidemic models. Some ensembles were unweighted, meaning each selected model contributed equally. Others were weighted according to their \( AIC_c \) rankings, for example, by combining the top 2 models (Ensemble 2) or the top 3 models (Ensemble 3), based on the ordering \( AIC_{c_1} \leq \ldots \leq AIC_{c_r} \), where \( i = 1, \ldots, r \).

The construction of these ensemble models follows the methodology detailed in \citep{chowell2022ensemble}. The 95\%  Prediction intervals (PIs) for the ensemble forecasts were computed using the same bootstrapping method described above, allowing us to quantify uncertainty in the ensemble outputs.

\subsection{Auto-regressive integrated moving average model}
The ARIMA model is a well-established statistical technique commonly used for forecasting time series data. It has been successfully applied by ~\citep{mondal2014study} in finance, by ~\citep{tektacs2010weather, dimri2020time} in weather prediction, and in infectious disease modeling by~\citep{chowell2022ensemble, long2023forecasting, iftikhar2023short, benvenuto2020application}. Due to its flexibility and simplicity, the ARIMA model has become a standard benchmark in epidemic forecasting~\citep{fattah2018forecasting, bleichrodt2024evaluating}.

An ARIMA model is defined by three components: the autoregressive (AR) part, which captures dependencies between current and past observations; the integrated (I) part, which applies differencing to stabilize the series; and the moving average (MA) part, which models the relationship between an observation and past forecast errors.

The general form of an ARIMA$(p,d,q)$ model is given by:
\begin{equation}
\phi(B)(1 - B)^d y_t = c + \theta(B)\epsilon_t,
\label{eq:arima}
\end{equation}
where $y_t$ is the observed time series at time $t$, $B$ is the backward shift operator such that $By_t = y_{t-1}$, and $\epsilon_t$ is a white noise error term. The polynomial $\phi(B) = 1 - \phi_1 B - \cdots - \phi_p B^p$ represents the autoregressive component, while $\theta(B) = 1 + \theta_1 B + \cdots + \theta_q B^q$ represents the moving average component. The differencing operator $(1 - B)^d$ is employed to transform the time series into a stationary form by removing trends.
In the special case where $d = 1$, the model operates on the first-differenced series $y_t' = y_t - y_{t-1}$. Higher-order differencing is used when required to remove more complex trends.

For this study, we fitted ARIMA models to weekly COVID-19 wastewater data. Model selection was performed automatically using the Hyndman-Khandakar algorithm \citep{hyndman2008automatic}, which selects optimal values of $p$, $d$, and $q$ based on information criteria. Forecasts were generated using an interactive R Shiny dashboard~\citep{bleichrodt2024statmodpredict}, and any negative predictions are set to zero.

\subsection{Generalized additive model}
The generalized additive model (GAM)  extends the generalized linear model by allowing smooth, non-linear functions of the predictors, enabling flexibility in capturing complex relationships between covariates and the response variable~\citep{wood2017generalized, baayen2020introduction}. This flexibility makes GAMs particularly useful for modeling time series data where trends may be complex and may not follow simple parametric forms.

In this study, we used GAMs to model weekly COVID-19 viral load with time as the sole predictor. Assuming a Gaussian error, the model is specified as follows:
\begin{equation}
y_t = \beta_0 + s(t) + \epsilon_t, \label{eq:gam}
\end{equation}
where \( \beta_0 \) is an intercept, \( s(t) \) is an unspecified smooth function of time, and \( \epsilon_t \sim N(0, \sigma^2) \) represents the modeling error at time $t$ ~\citep{bleichrodt2024statmodpredict, Shafi}. Although we adopt the Gaussian family here, the GAM framework readily accommodates alternative distributions if the normality assumption is violated.

The smooth function \( s(t) \) is represented as a linear combination of basis functions:
\begin{equation}
s(t) = \sum_{k=1}^{K} \beta_k b_k(t), \label{eq:basis}
\end{equation}
where \( \{b_k(t)\} \) are predefined basis functions and \( \{\beta_k\} \) are the coefficients estimated from the data. In our implementation, we used the \texttt{gam()} function from the \texttt{mgcv} package in R~\citep{wood_gam}, which by default employs penalized regression splines (e.g., thin plate splines or cubic splines). All the forecasts for this model were generated on the R-shiny dashboard \citep{bleichrodt2024statmodpredict}.

The number of basis functions \( K \) was selected based on the length of the calibration window for each forecast period. To avoid overfitting and control the level of smoothness, a penalty was applied to the coefficients\( \{\beta_k\} \), and the optimal smoothness parameter was selected via generalized cross-validation~\citep{wood2017generalized, bleichrodt2024statmodpredict}.

GAMs offer several advantages for epidemic forecasting, including flexibility to model non-linear dynamics, interpretability through additive components, and the capacity to incorporate domain knowledge via the choice of smooth terms ~\citep{Shafi, wood2017generalized}. Although GAMs capture short-term patterns effectively, their utility diminishes when extrapolating beyond the calibration window. Because the spline basis functions are fitted to the calibration data, the model quickly loses predictive validity once the series extends past that window, causing forecast accuracy to decline after only a few time steps (in our case, one to two weeks). Consequently, we use the GAM solely as a near-term benchmark whose forecasts complement, rather than replace, those from mechanistic or ensemble models.

\subsection{Simple Linear Regression (SLR)}
Simple Linear Regression is a foundational statistical method that models the relationship between a response variable and a single predictor using a linear equation. For our time series forecasting application, we model the WVAL as a linear function of time:
\begin{equation}
{y_t = \beta_0 + \beta_1 t + \epsilon_t,}
\end{equation}
where $y_t$ is the WVAL at time $t$, $\beta_0$ is the intercept, $\beta_1$ is the slope coefficient representing the linear trend, and $\epsilon_t \sim N(0, \sigma^2)$ is the error term. The parameters $\beta_0$ and $\beta_1$ are estimated using ordinary least squares, minimizing the sum of squared residuals. While SLR provides a simple baseline for comparison and can capture overall trends, its assumption of a constant linear relationship limits its ability to model the complex, non-linear dynamics often observed in epidemic time series. Any negative predictions from the SLR model were set to zero. Despite its limitations, including SLR in our analysis provides a useful benchmark to assess the relative performance gains achieved by more sophisticated modeling approaches.

\subsection{Meta's (Facebook's) Prophet model}

Originally introduced for business analytics, Facebook's Prophet model \citep{taylor2018forecasting} has become increasingly popular across different fields, including epidemiology. It has been applied to model and forecast disease trajectories such as those observed in COVID-19 \citep{satrio2021time, battineni2020forecasting, kirpich2022excess} and mpox outbreaks \citep{long2023forecasting}. Prophet is built on the assumption that the underlying time series \( y(t) \) can be decomposed into interpretable components representing different temporal patterns:
\begin{equation}
    y(t) = g(t) + s(t) + h(t) + \epsilon_t, 
\end{equation}

where each component captures different aspects of the time series:

\begin{itemize}
    \item $g(t)$: The trend component that captures long-term, non-seasonal changes in the time series
    \item $s(t)$: The seasonal component that models recurring patterns (e.g., weekly, monthly, or yearly cycles)
    \item $h(t)$: The holiday/event component that accounts for the impact of special events or anomalies
    \item $\epsilon_t$: The error term representing random fluctuations not captured by the other components
\end{itemize}

For our analysis, we used the default settings of the R \texttt{prophet} function from the `prophet' package, as described by Taylor \citep{taylor_prophet}. Forecasting was conducted using the \texttt{predict} function from the same package \citep{stats_predict}. Any negative forecasted values were set to zero. Additional details on the model fitting process are available in \citep{bleichrodt2024statmodpredict, taylor_prophet}.

\subsection{Model Nomenclature and Forecasting Framework}
\label{sec:nomenclature}

The $n$-sub-epidemic framework generates multiple candidate models at each forecast date by fitting models with $n=1$ and $n=2$ sub-epidemics to the calibration data (we set $n_{\max}=2$). We rank these candidates using corrected Akaike Information Criterion (AIC$_c$) and select the top three performers. Model names reference this ranking:

\begin{itemize}
\item \textbf{Rank 1}: Best-performing \textit{n}-sub-epidemic model (lowest AIC$_c$)
\item \textbf{Rank 2}: Second-best \textit{n}-sub-epidemic model
\item \textbf{Rank 3}: Third-best \textit{n}-sub-epidemic  model
\end{itemize}

Ensemble models combine these ranked models using two weighting approaches:
\begin{itemize}
\item \textbf{EM2 W / EM3 W}: Weighted ensembles of the top 2 or 3 models, with weights derived from AIC$_c$ differences using Akaike weights (Burnham \& Anderson, 2002). Specifically, model $i$ receives weight $w_i = \exp(-\Delta_i/2) / \sum_j \exp(-\Delta_j/2)$, where $\Delta_i = \text{AIC}_{c,i} - \min(\text{AIC}_c)$.
\item \textbf{EM2 UW / EM3 UW}: Unweighted ensembles where each of the top 2 or 3 models contributes equally ($w_i = 1/k$). This approach provides robustness when model rankings are unstable or AIC$_c$ differences are negligible.
\end{itemize}

Comparison models (ARIMA, GAM, SLR, Prophet) maintain their conventional names and are fitted using the procedures described in preceding subsections. 
All forecasts use epidemiological weeks (epiweeks) as defined by the CDC's MMWR calendar, where each week runs Saturday through Friday. This aligns with NWSS reporting structure and ensures temporal consistency.

For a forecast generated at the end of epiweek $t$, the targets are:
\begin{itemize}
\item 1-week ahead: epiweek $t+1$ (7 days)
\item 2-week ahead: epiweek $t+2$ (14 days)
\item 3-week ahead: epiweek $t+3$ (21 days)
\item 4-week ahead: epiweek $t+4$ (28 days)
\end{itemize}

We generated retrospective out-of-sample forecasts using a rolling, fixed-length
calibration window. At each forecast origin $t$ (March 2022--September 2024), we:
\begin{enumerate}
  \item Fit all models using only the \textbf{most recent 10 weeks of data}, i.e., $[t-9, t]$;
  \item Generate 1--4 week-ahead forecasts;
  \item Use the bootstrap method to construct the 95\% PIs and get the medians;
  \item Advance one week and repeat.
\end{enumerate}

This protocol ensures no future information leaks into training data. Model parameters are re-estimated at each forecast time, allowing the models to adapt to evolving dynamics. The resulting forecast-observation pairs reflect operational deployment conditions where public health agencies would use only past data to predict future trends.

\section{Model Evaluation}
In this study, we assessed the forecasting performance of each model by computing the average values of four key metrics: Mean Absolute Error (MAE), Mean Squared Error (MSE), Weighted Interval Score (WIS), and 95\% Prediction Interval (PI) coverage. Here, the term "average" denotes the mean of each metric, calculated across all forecasting periods for every combination of model, location, and forecast horizon.

\subsection{Mean Absolute Error (MAE) }
The mean absolute error represents the average magnitude of errors in the model predictions \citep{willmott2005advantages,kaundal2006machine}. It is calculated as the average difference between the predicted values and the actual values in the test dataset, with equal weight assigned to all individual differences. The range of MAE is from 0 to infinity, where smaller values indicate better model performance. Due to this property, it is sometimes referred to as a negatively oriented score \citep{baran2016censored}.
\begin{equation}
 \text{MAE} = \frac{1}{N} \sum_{j=1}^N \lvert y_j - \hat{y}_j \rvert.  
\end{equation}

\subsection{Mean Square Error (MSE)}
The mean square error is another widely used metric for evaluating the performance of regression models \citep{willmott2005advantages}. It computes the squared differences between the predicted and actual values, summing them up and averaging over all data points. Squaring the differences removes negative signs and gives greater emphasis to larger errors. A lower MSE indicates a better fit to the data. The formula for MSE is given by:
\begin{equation}
   \text{MSE} = \frac{1}{N} \sum_{j=1}^N (y_j - \hat{y}_j)^2. 
\end{equation}

\subsection{Weighted Interval Score (WIS)}
The Weighted Interval Score quantifies how closely a probabilistic forecast distribution aligns with the observed outcome, using the scale of the data. It combines the absolute error of the median forecast with a weighted sum of interval scores at different prediction levels, providing a comprehensive measure of forecast accuracy \citep{cramer2022evaluation}.

WIS was introduced by \citep{bracher2021evaluating} as an effective scoring rule for probabilistic forecasting of epidemics. For a forecast with prediction intervals at nominal coverage levels
\begin{equation*}
\alpha \in \{0.02, 0.05, 0.10, \ldots, 0.90, 0.95, 0.98\}~, 
\end{equation*}
the Interval Score (IS) at level $\alpha$ is defined as:
\begin{equation}
\text{IS}_\alpha(F, y) = (u_\alpha - l_\alpha) + \frac{2}{\alpha}(l_\alpha - y)\mathbb{1}\{y < l_\alpha\} + \frac{2}{\alpha}(y - u_\alpha)\mathbb{1}\{y > u_\alpha\}
\end{equation}
where $l_\alpha$ and $u_\alpha$ are the lower and upper bounds of the $(1-\alpha) \times 100\%$ prediction interval, and $y$ is the observed value. The first term rewards narrow intervals (sharpness), while the penalty terms penalize under-coverage when observations fall outside the interval.

The Weighted Interval Score aggregates these interval scores:
\begin{equation}
\text{WIS}_{\alpha_{0:K}}(F, y) = \frac{1}{K + \frac{1}{2}} \left( w_0 |y - m| + \sum_{k=1}^{K} w_k IS_{\alpha_k}(F, y) \right),
\end{equation}
where $w_k = \frac{\alpha_k}{2}$ for $k = 1, 2, \ldots, K$ , $m$ is the median (50th percentile) of the forecast distribution F, \(\alpha_k\) (for \(k = 1,\dots,K\)) is the nominal coverage level of the central \((1 - \alpha_k)\times 100\%\) prediction interval and $w_0 = \frac{1}{2}$. Lower WIS indicates better forecast performance. Because WIS inherits the units of the forecast target (log$_{10}$ WVAL in our case), differences in WIS values can be interpreted directly in terms of forecast error magnitude.

\subsection{Coverage rate of 95\% Prediction Interval (95\% PI)}
The 95\% PI is a range within which future observations are expected to fall with a 95\% probability, based on the predictive distribution \citep{montgomery2015introduction}. It provides a measure of the uncertainty in a forecast and is crucial for assessing the reliability of predictive models. A well-calibrated model should produce prediction intervals that closely match the observed proportion of actual values falling within the interval.

The coverage rate of the 95\% PI can be expressed as:
\begin{equation}
 \frac{1}{N} \sum_{j=1}^{N} \mathbf{1} \{ L_{j} < Y_{j} < U_{j} \},
\end{equation}
where $L_{j}$ and $U_{j}$ represent the lower and upper bounds of the 95\% prediction interval, respectively. The variable $Y_{j}$ denotes the observed value, and $\mathbf{1}\{\cdot\}$ is an indicator function that equals 1 if $Y_{j}$ falls within the prediction interval, and 0 otherwise \citep{chowell2024subepipredict}.

\subsection{Baseline Model and Forecast Skill Scores}
\label{sec:baseline}

To contextualize model performance improvements, we use Simple Linear Regression (SLR, described in Section 2.7) as our baseline reference model. SLR provides a minimal-skill benchmark representing simple trend extrapolation without capturing complex epidemic dynamics.

We compute forecast skill scores to quantify improvement over this baseline by comparing the \emph{average} error across all forecast dates. For example, for MAE, we define
\begin{equation}
\text{Skill Score} = 1 - \frac{\overline{\text{MAE}}_{\text{model}}}{\overline{\text{MAE}}_{\text{SLR}}},
\end{equation}
where $\overline{\text{MAE}}_{\text{model}}$ and $\overline{\text{MAE}}_{\text{SLR}}$ denote the mean MAE across all forecast dates for a given model and for the SLR baseline, respectively. Positive values indicate that the model outperforms SLR (e.g., $\text{SS} = 0.40$ implies a 40\% reduction in error compared to linear trend extrapolation). Skill scores provide a scale-independent performance metric across regions with different baseline error magnitudes. Skill scores for MAE, MSE, and WIS across all models, regions, and horizons are presented in Supplementary Figures S6--S10 and referenced in the Results where appropriate.

\section{Results}

Using a rolling-origin retrospective evaluation design, we generated 133 out-of-sample weekly forecast sets spanning March 5, 2022, through September 14, 2024, producing over 10,000 forecast-observation pairs across all regions, models, and forecast horizons. At each forecast origin, models were fit only to the preceding 10 weeks of WVAL data and then evaluated against subsequently observed values at 1- to 4-week horizons. Figures \ref{fig:4}, Supplementary Figures~S2, S4, S6, and S8 present aggregated performance metrics (mean values computed across all 133 forecast weeks) to identify systematic patterns in model performance. To complement these averages, we also present the full distributions of MAE, MSE, and WIS across forecast weeks in Supplementary Figures S1--S5 (boxplots for each region), which highlight variability and consistency of model performance over time. This aggregation approach balances comprehensive evaluation with interpretability. Throughout the Results, we reference models using the nomenclature established in Section \ref{sec:nomenclature}: Rank 1-3 (top n-sub-epidemic models by AIC$_c$), EM2 W/UW and EM3 W/UW (2- and 3-model ensembles, weighted and unweighted), and comparison models ARIMA, GAM, SLR, and Prophet.

\begin{figure} [H]
    \centering
    \includegraphics[width=.8\textwidth]{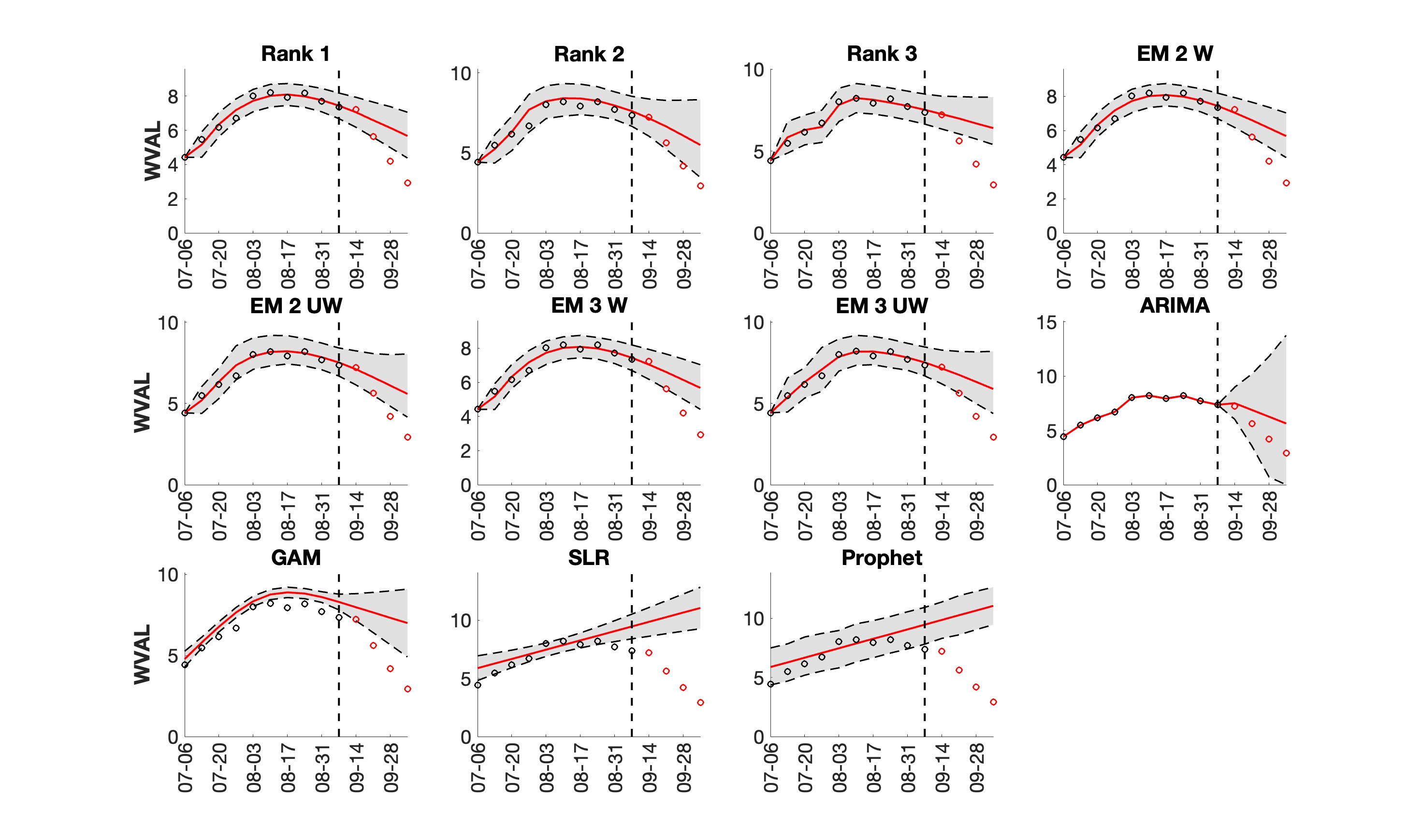}
    \caption{\footnotesize{{Forecasts of WVAL at the national level for an example forecast date (originating on September 7, 2024) across all models. Each panel shows the median prediction (red solid line) and the corresponding 95\% prediction interval (gray shaded region bounded by black dashed lines). Black circles denote calibration data, while red circles indicate observations used for evaluation. The vertical dashed line marks the end of the calibration period. This example illustrates the varying widths of prediction intervals across models, with ensemble approaches generally providing wider but more reliable uncertainty bounds than single models such as SLR or Prophet.}}}
    \label{figure:2}
\end{figure}

\begin{figure}[H]
    \centering
    \includegraphics[width=.8\textwidth]{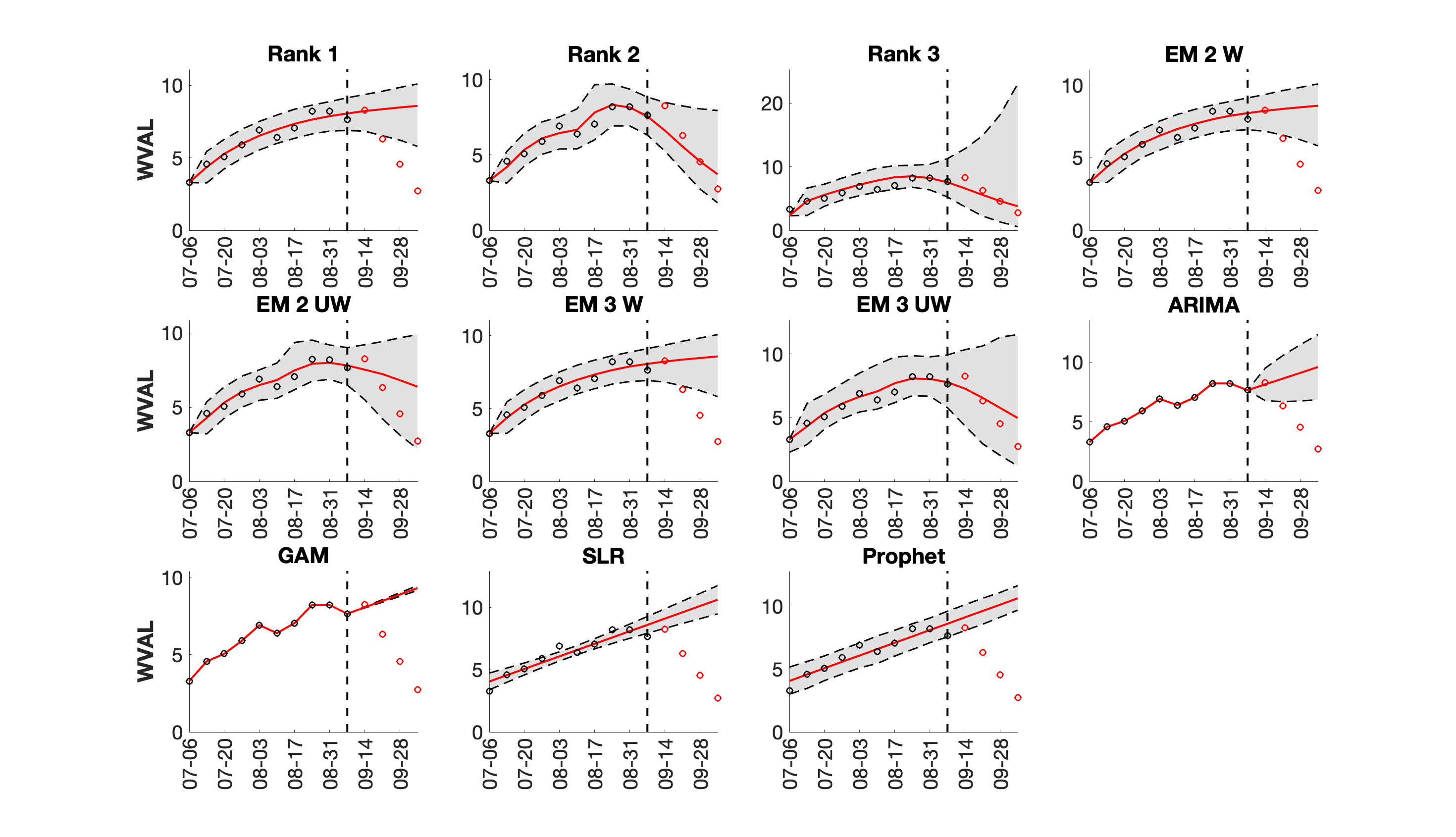}
    \caption{\footnotesize{{Forecasts of WVAL for the Midwest region for an example forecast date (originating on September 7, 2024) across all models. Each panel shows the median prediction (red solid line) and the corresponding 95\% prediction interval (gray shaded region bounded by black dashed lines). Black circles denote calibration data, while red circles indicate observations used for evaluation; the vertical dashed line marks the end of the calibration period. The GAM model exhibits particularly tight prediction intervals during calibration, making its gray region difficult to distinguish from the median curve. The sub-epidemic models more accurately track the observed declining trend, whereas several statistical models tend to overestimate future WVAL values.
} }}
    \label{figure:3}
\end{figure}

We start the Results section by highlighting forecasts from September 7, 2024, one of the final evaluation dates in our retrospective study, to show how different models behave when predicting beyond the calibration period. Figure~\ref{figure:2} displays national-level forecasts from all models, while Figure~\ref{figure:3} shows the same for the Midwest region. In both cases, sub-epidemic models closely follow the observed decline in WVAL and maintain prediction intervals that are wide enough to include future observations. In contrast, statistical models like SLR and Prophet tend to produce much narrower intervals that quickly diverge from the actual data. Together, these plots preview two key patterns we explore in the following sections: the strong performance of ensemble models at 3--4 week horizons and their ability to better anticipate regional turning points.

\subsection{National}
\begin{figure}[H]
    \centering
    \includegraphics[width=.75\textwidth]{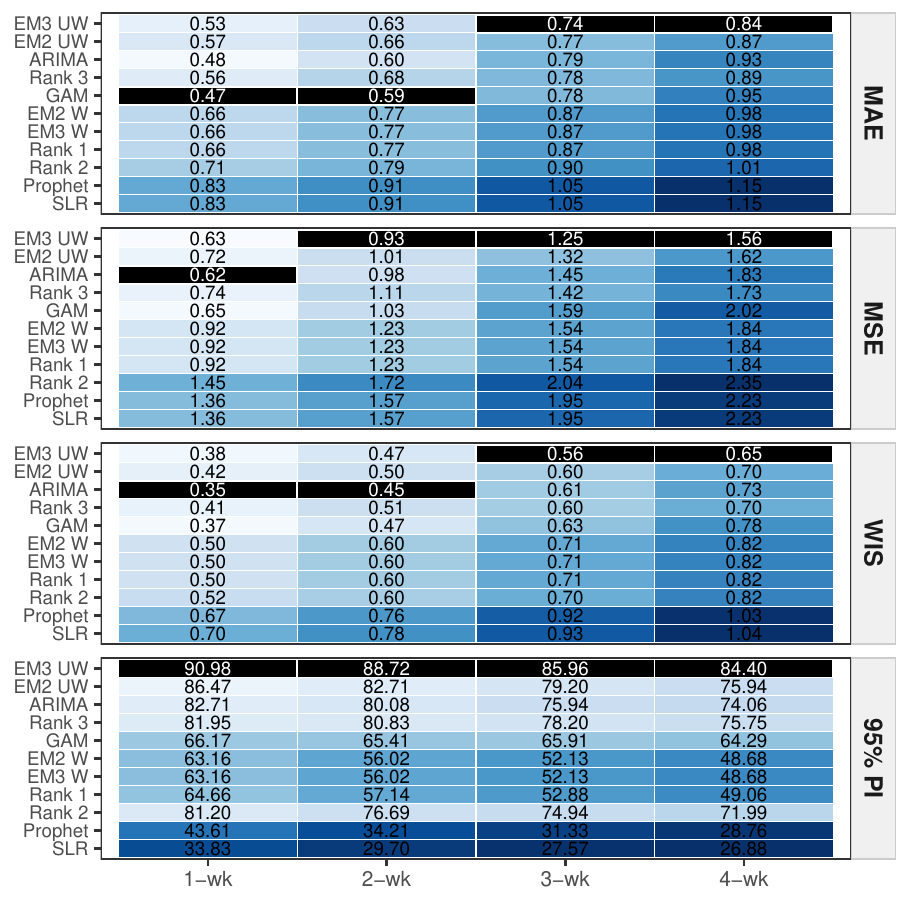}
    \caption{\footnotesize{Log-transformed averages of MAE, MSE, and WIS, and the (untransformed) average 95\% PI coverage across all models for 1--4 week horizons in the National region, spanning the period from March 5, 2022, to September 14, 2024 (133 forecasts). 
Lighter shades indicate smaller (better) values for MAE, MSE, and WIS, whereas darker shades indicate larger errors. For coverage, values closer to 95\% indicate better performance. 
Black cells highlight the best-performing model(s) for each forecast horizon and metric.
}}
    \label{fig:4}
\end{figure}

Results for the National region are summarized in Fig.~\ref{fig:4}, which pools all metrics 
(MAE, MSE, WIS, Coverage95) across 1--4-week horizons into a single panel. 
Errors increase with forecast horizon, but relative model rankings remain stable. 
 
At the 1-week horizon, ARIMA is among the best-performing models, achieving the lowest MSE (0.65) and WIS (0.35). GAM also performs strongly, with MAE 0.47, MSE 0.65, and WIS 0.37. Several \textit{n}-sub-epidemic models (EM2 UW and EM3 UW) provide similarly competitive results, with WIS values of  0.42 and 0.38, respectively. For uncertainty quantification, EM3 UW attains the highest 95\% PI coverage (90.98\%), followed closely by EM2 UW (86.47\%).

At the 2-week horizon, GAM again is part of the set of models that provided the lowest error metrics, achieving MAE 0.59, MSE 1.03, and WIS 0.47. ARIMA performs similarly, with MAE 0.60, MSE 0.98, and WIS 0.45. EM3 UW and EM2 UW produce MAE values of 0.63 and 0.66 and WIS values of 0.47 and 0.50, 
remaining competitive. EM3 UW again yields the highest 95\% PI coverage at this horizon (88.72\%).

At the 3-week horizon, EM3 UW emerges as the strongest overall model, achieving the lowest 
MAE (0.74), MSE (1.25), and WIS (0.56). Rank3 and EM2 UW also perform well, with MAE 0.78 \& 0.77 respectively and the same WIS value of 0.60. ARIMA also provided competitive values at this horizon (MAE 0.78, MSE 1.45, WIS 0.61). EM3 UW again attains the highest 95\% PI coverage (85.96\%).

At the 4-week horizon, EM3 UW continued to be among the strongest models, achieving the lowest WIS (0.65) and one of the lowest MAE values (0.84). EM2 UW also performs well, with MAE 0.87, MSE 1.62, and WIS 0.70. For uncertainty quantification, EM3 UW again provides the highest 95\% PI coverage (84.40\%), indicating consistently well-calibrated forecast intervals at the national scale.

Overall, these results indicate that GAM and ARIMA offer highly accurate 1--2-week-ahead forecasts for the National region, while EM2 UW and EM3 UW provide the most reliable 3--4-week-ahead forecasts. Some models achieve very similar accuracy at certain horizons--for example, at the 1-week horizon, GAM and ARIMA produced comparable MAE values (0.47 and 0.48).

To help evaluate the magnitude and variability of these differences, we present the distributions of all metrics across models, forecast periods, and horizons at the national level in Supplementary Fig.~S1 (with corresponding regional distributions in Supplementary Figs.~S3, S5, S7, and S9). Performance metric distributions across all 133 forecast periods reveal consistency of model rankings. Focusing on the national level, Supplementary Fig.~S1 shows that GAM and EM3 UW consistently achieve the lowest MAE values among all models, with their distributions clearly separated from SLR and Prophet. The boxplots show partial overlap among the top-performing models (GAM, EM3 UW, ARIMA, Rank 3), indicating that while ensemble approaches offer reliable advantages, performance differences among leading models vary across forecast periods. This pattern suggests that the ensemble approach provides dependable, though not universal, improvements across changing epidemic dynamics.

Skill scores relative to the SLR baseline (Section \ref{sec:baseline}) quantify forecast improvement beyond simple trend extrapolation. At the national level, EM3 UW achieves MAE skill scores of 46\% at 1-week ahead and 40--41\% at 2--4 week horizons, representing substantial error reduction compared to linear trend forecasting (Supplementary Figure S~10). For MSE, skill scores reach 55--70\%, while WIS skill scores range from 49--54\% across all horizons. These consistently strong values demonstrate that ensemble forecasting provides meaningful improvements over naive baseline methods throughout the forecast window. Regional skill score patterns (Supplementary Figures S11–S14) show similar results, with ensemble models consistently achieving positive skill scores across all regions and horizons.

\subsection{Midwest}

Results for the Midwest region are summarized in supplementary Fig. 2. Forecast errors increase with horizon length, but overall model rankings remain relatively stable. At the 1-week horizon, GAM and ARIMA are among the best-performing models, with GAM attaining the lowest MAE (0.63) and the lowest MSE (0.96), while ARIMA also performs competitively, particularly for MAE (0.64) and WIS (0.47). The EM3 UW ensemble provides the highest 95\% PI coverage (87.97\%), indicating strong uncertainty quantification even for near-term forecasts.

At the 2-week horizon, ARIMA and GAM are again among the best-performing models. 
ARIMA attained the lowest MAE (0.71), MSE (1.15), and WIS (0.54), while GAM provided a very competitive MAE (0.73),  MSE (1.27) and WIS (0.57) values. EM3 UW also performed with providing very close MAE, MSE, and WIS values.
EM3 UW again attains the highest 95\% PI coverage (85.71\%), indicating a well-calibrated ensemble option at shorter horizons.

Performance patterns shift at horizons 3 and 4. At the 3-week horizon, EM3 UW is part of the best-performing set of models, with the lowest MAE (0.91), the lowest MSE (1.65), and the lowest WIS (0.70), while also maintaining high 95\% PI coverage (83.46\%). At the 4-week horizon, EM3 UW continues to be among the top-performing models in terms of MAE (1.03), MSE (1.98), WIS (0.81), and PI coverage (81.58\%). The EM2 UW also provided very similar values compared to EM3 UW at horizons 3 and 4.

Overall, these patterns suggest that GAM and ARIMA are well suited for 1--2--week forecasting in the Midwest, whereas the unweighted ensembles, specifically EM3 UW and EM2 UW, are among the most reliable options for 3--4 weeks, balancing accuracy and uncertainty quantification. Several models exhibit very similar performance in this region; for example, at the 3-week horizon, EM2 UW had an average WIS of 0.71 compared with 0.70 for EM3 UW. To help assess the magnitude of these small differences, we present the distributions of the metrics across all models, forecast periods, and horizons in Supplementary Fig.~S3, and the corresponding skill scores relative to the SLR baseline in Supplementary Fig.~S11, rather than relying solely on point estimates.

\subsection{Northeast}
Supplementary Fig.~S4 summarizes the results for the Northeast region.  
Across all horizons, this region shows some of the highest MAE, MSE, and WIS values, indicating that the Northeast was among the most challenging settings for forecasting in our study. We discuss potential reasons for this in detail in Section~5.1. 

At the 1-week horizon, ARIMA and EM3 UW are the leading models. 
EM3 UW attains the lowest MAE (0.80), MSE (1.33), and WIS (0.59), with ARIMA, Rank2, and EM2 UW yielding very similar WIS values (0.61, 0.62, and 0.61, respectively). Both EM3 UW and the Rank3 of the \textit{n}-sub-epidemic model attain the highest 95\% PI coverage (86.47\%), indicating well-calibrated uncertainty at this horizon.

At the 2-week horizon, ARIMA is clearly the best-performing model on error-based metrics, achieving the lowest MAE (0.77), MSE (1.28), and WIS (0.58). 
EM3 UW remains competitive, with slightly larger error values (MAE 0.85, MSE 1.46, WIS 0.63), 
and again attained high 95\% PI coverage (86.09\%), comparable to the Rank3 model (86.47\%).

Performance patterns changed at horizons 3 and 4. 
At the 3-week horizon, EM3 UW provided the lowest MAE (0.92), MSE (1.66), 
and WIS (0.69) among all models. It also delivered high 95\% PI coverage (85.96\%), only slightly below the Rank3 model (86.47\%). ARIMA remains reasonably competitive but with higher errors (MAE 1.41, MSE 2.83, WIS 1.00).

At the 4-week horizon, EM3 UW is the clear top performer, achieving the lowest MAE (1.00), 
MSE (1.95), and WIS (0.77), while also providing the highest 95\% PI coverage (85.15\%). 
The Rank~2 and EM2--UW ensembles are the next-best models on the error metrics at this horizon, with MAE values of 1.03, MSE values of 2.13 and 2.05, and WIS values of 0.82 and 0.80, respectively, illustrating the strength of unweighted ensemble methods for 3- and 4-week-ahead forecasts in this region.

Overall, these results indicate that ARIMA and the unweighted ensemble models offer the strongest 1--2-week-ahead performance in the Northeast, whereas unweighted \textit{n}-sub-epidemic ensembles, particularly EM3 UW, provide the most reliable 3--4-week-ahead performance. 
In this region, some models yield very similar error values; for example, at the 1-week horizon, the EM3 UW model achieves an MAE of 0.80 compared with 0.82 for ARIMA. 
To help assess the magnitude of these differences, we present the distributions of all metrics across models, forecast periods, and horizons in Supplementary Fig.~S5, and the corresponding skill scores relative to the SLR baseline in Supplementary Fig.~S12, rather than relying solely on point estimates.

\subsection{South}
Supplementary Fig.~S6 summarizes the results for the South region. 
This region exhibits moderate error levels relative to the Northeast and Midwest, with clearer 
separation among models and more stable performance patterns across horizons. 
Overall, ARIMA and EM3 UW consistently rank among the best-performing models across all forecast horizons, with GAM also performing well at 1--2 weeks.

At the 1-week horizon, ARIMA is the strongest model on error-based metrics, achieving the lowest MAE (0.56), MSE (0.80), and WIS (0.42). 
GAM and several \textit{n}-sub-epidemic models also perform well, including EM3 UW and Rank3, with MAE values of 0.58 and 0.63 and WIS values of 0.43 and 0.48, respectively. 
GAM attains MAE 0.59, MSE 0.88, and WIS 0.46. In terms of uncertainty quantification, EM3 UW attains the highest 95\% PI coverage (90.23\%), closely followed by EM2 UW (88.72\%).

At the 2-week horizon, ARIMA again provided the lowest error values, with MAE 0.74, 
MSE 1.34, and a very competitive WIS of 0.58 compared with EM3 UW (0.57). 
GAM and the unweighted ensembles remain competitive: GAM attains MAE 0.78, MSE 1.54, and WIS 0.64, while EM3 UW and EM2 UW produce similar MAE values (0.75 and 0.78) and strong predictive interval coverage (87.22\% and 82.33\%, respectively), highlighting their consistency at this horizon.

At the 3- and 4-week horizons, ARIMA continues to provide the lowest error values. 
At the 3-week horizon, ARIMA attains MAE 0.86, MSE 1.68, and WIS 0.68. 
At the same horizon, EM3 UW and EM2 UW yield very competitive results, with MAE 0.89 and 0.90, MSE 1.88 and 1.84, and WIS 0.73 and 0.81, respectively. 
At the 4-week horizon, ARIMA remains the top performer, providing the lowest MAE (0.98), 
MSE (1.99), and WIS (0.79), while EM3 UW remains a strong competitor with similar values.

Across all horizons, EM3 UW attains the highest 95\% PI coverage, indicating well-calibrated 
forecast uncertainty in this region.

Overall, these results indicate that ARIMA, GAM, EM2 UW, and EM3 UW provide the strongest 
1--2-week-ahead performance for the South region, while ARIMA and EM3 UW remain the best-performing models for 3--4-week-ahead forecasts. 
In this region, some models produce very similar metric values; for example, at the 2-week horizon, ARIMA yields a WIS of 0.58 compared with 0.57 for EM3 UW. 
To help assess the magnitude and variability of these differences, we present the distributions of all metrics across models, forecast periods, and horizons in Supplementary Fig.~S7, and the corresponding skill scores relative to the SLR baseline in Supplementary Fig.~S13, rather than relying solely on point estimates.

\subsection{West}
Supplementary Fig.~S8 presents the results for the West region. There is clear separation among models and strong performance from both ARIMA and the unweighted \textit{n}-sub-epidemic ensembles, particularly EM3--UW.  Overall, these models consistently rank among the top-performing approaches across all forecast horizons.

At the 1-week horizon, EM3 UW and EM2 UW are the leading models based on metrics. 
EM3 UW achieved the lowest MAE (0.57), the lowest MSE (0.76), and the lowest WIS (0.42). 
EM2 UW performed similarly, with MAE 0.60, MSE 0.83, and WIS 0.43. GAM also performed well, providing MAE 0.59, MSE 0.86, and WIS 0.47. For uncertainty quantification, EM3 UW yields exceptionally high 95\% PI coverage (96.24\%), the highest of any model in this region and significantly above the other regions.

At the 2-week horizon, EM3 UW remains the best-performing model on all three error metrics, 
achieving the lowest MAE (0.70), MSE (1.17), and WIS (0.52). EM2 UW again performs competitively, with MAE 0.72, MSE 1.18, and WIS 0.55. ARIMA provided MAE 0.79, MSE 1.37, and WIS 0.62, remaining competitive but with slightly larger errors. EM3 UW continued to provide the highest 95\% PI coverage (94.36\%) at this horizon.

At the 3-week horizon, EM3 UW again is part of the best-performing models, with one of the lowest MAEs (0.84), a competitive MSE (1.62) compared to EM2 UW (1.51) and the lowest WIS (0.67). Most of the sub-epidemic models provided similar values across metrics. EM3 UW further provides the highest 95\% PI coverage (90.48\%) at this horizon.

At the 4-week horizon, EM3 UW continues to be among the top performers, achieving competitive 
MAE (0.95) and MSE (1.97), compared to EM2 UW MAE (0.94) and MSE (1.84). The Rank 1 model also performs competitively with MAE 0.99, MSE 1.90, and WIS 0.80. For this horizon, EM3 UW also attains the highest 95\% PI coverage (87.78\%), highlighting its consistently strong calibration in this region.

Overall, these results show that the West region benefits from strong performance by both ARIMA, GAM, and the unweighted \textit{n}-sub-epidemic ensembles across a 1--2 week forecast. The \textit{n}-sub-epidemic models provide the most reliable performance across all horizons, achieving the lowest error metrics and the highest PI coverage values. Clearly several models produce similar results at some horizons; for example, at the 3-week horizon, GAM and ARIMA yielded MAE values of 0.91 and 0.92, respectively. To help assess the magnitude and variability of these differences, we present the distributions of all metrics across models, forecast periods, and horizons in Supplementary Fig.~S9, and the corresponding skill scores relative to the SLR baseline in Supplementary Fig.~S14, rather than relying solely on point estimates.

\begin{figure}[H]
    \centering
    \includegraphics[width=.8\textwidth]{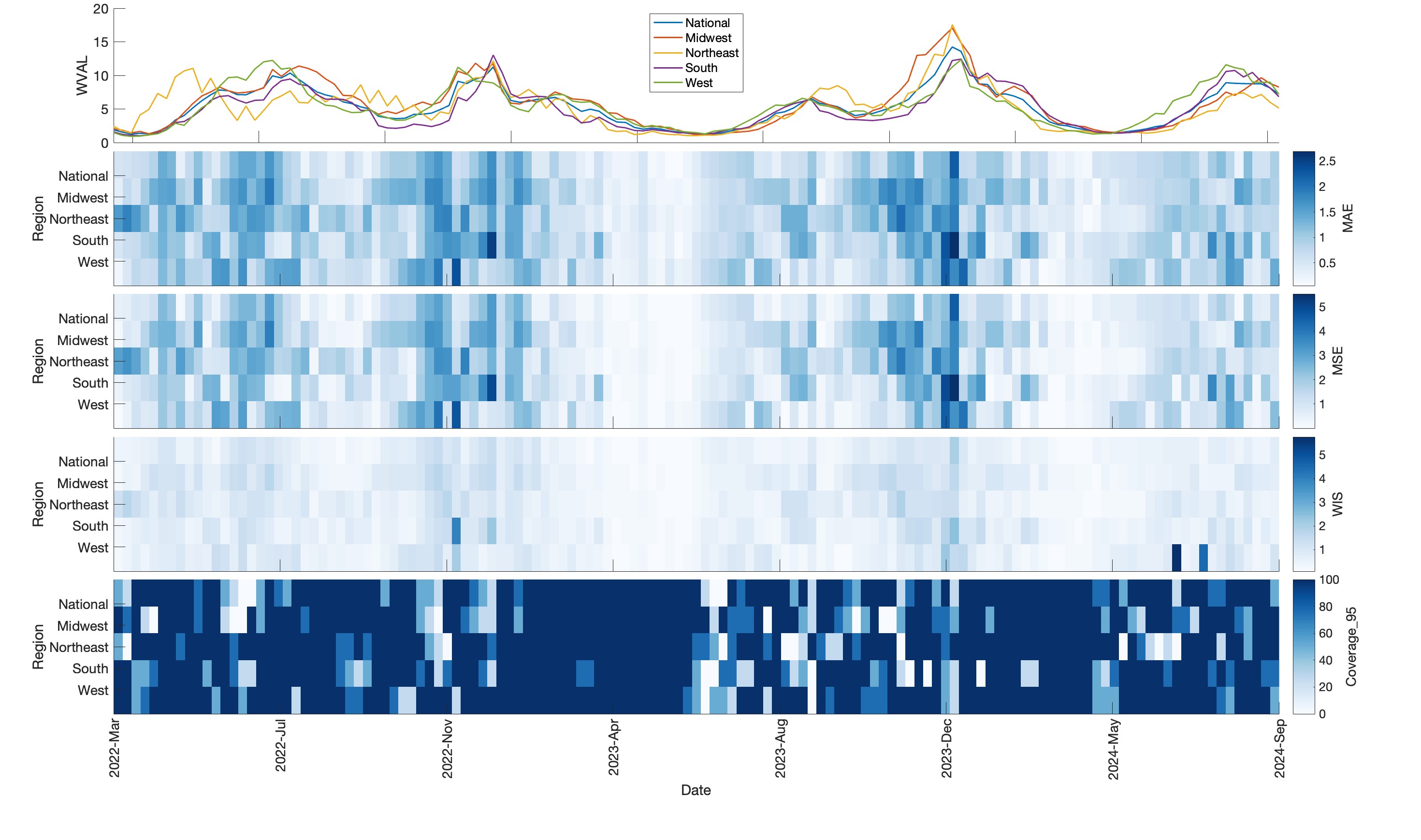}
    \caption{\footnotesize {Temporal dynamics of forecast performance for EM3 UW model. Top panel: Weekly viral activity levels (WVAL, log$_{10}$ scale) across five U.S. regions from March 2022 to September 2024. Lower panels: Heatmaps showing temporal variation in four performance metrics (MAE, MSE, WIS, and 95\% PI coverage) for 4-week ahead forecasts. Darker blue in error metric panels indicates higher errors. The coverage panel uses inverted coloring (darker = higher coverage = better performance). Error magnitudes increase during high WVAL periods, but this is partly expected because both WVAL and errors are on the same log scale; higher baseline values naturally yield larger absolute errors. The key scientific question is whether errors increase \textit{disproportionately} during surges, indicating true model degradation during critical periods. Future analysis should examine skill scores (errors relative to baseline) across different WVAL regimes to isolate genuine performance degradation from scale-dependent error increases.}}
    \label{fig:8}
\end{figure}

\begin{figure}[H]
    \centering
    \includegraphics[width=.8\textwidth]{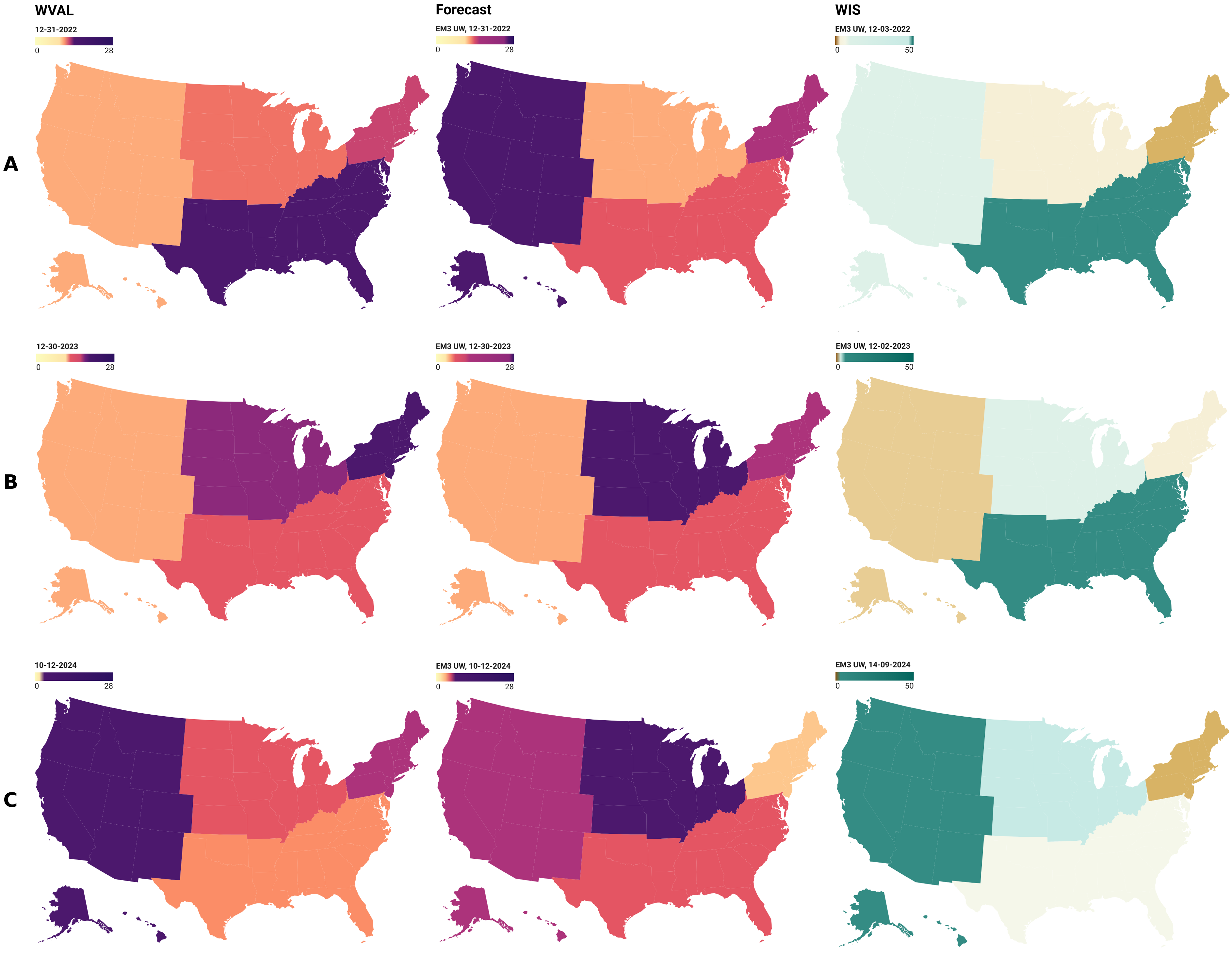}
    \caption{\footnotesize {The figure displays the WVAL values and model performance for three different forecasting periods, each with a 4-week forecast horizon: (A) a forecast conducted on 12-03-2022 targeting the week of 12-31-2022; (B) a forecast conducted on 12-02-2023 targeting the week of 12-30-2023; and (C) a forecast conducted on 09-14-2024 targeting the week of 10-12-2024.
Each row corresponds to one of these forecast periods. The first column shows the actual WVAL values observed during the target week. The second column presents the 4-week-ahead predictions from the EM3 UW model, and the third column displays the associated WIS. 
This figure highlights how the model performs across different periods and regions. The spatial heterogeneity evident in these maps underscores the importance of region-specific modeling approaches, as viral activity patterns and forecast accuracy vary substantially across geographic areas.
}}
    \label{fig:9}
\end{figure}

\section{Discussion}
This study highlights the effectiveness of retrospective forecasting using wastewater surveillance data to track COVID-19 trends across U.S. regions. By evaluating multiple models across various forecast horizons, we identified key differences in predictive accuracy and uncertainty quantification. The unweighted ensemble models within the \textit{n}-sub-epidemic framework, especially EM3 UW and EM2 UW, stood out for their reliability in forecasting 3- to 4-week ahead. On the other hand, statistical models like ARIMA and GAM were better suited for 1- to 2-week forecasts. These findings reinforce the advantage of combining overlapping sub-epidemics to capture complex transmission dynamics, especially in settings characterized by multiple peaks and extended plateaus in wastewater signals \citep{chowell2022ensemble}.

Our results align with previous research showing that ensemble models improve epidemic forecast performance by balancing the strengths and weaknesses of individual models \citep{cramer2022evaluation,yamana2017individual}. Specifically, EM3 UW consistently outperformed other models across metrics such as MAE, WIS, and 95\% prediction interval (PI) coverage, while EM2 UW in some cases minimized large prediction errors (MSE). This supports the use of ensemble modeling to enhance both accuracy and robustness in epidemic forecasting \citep{cramer2022evaluation}.
The robustness of ensemble approaches is further supported by large-scale international validation: analysis of 48 forecasting models across 32 European countries over 52 weeks found that 
ensembles outperformed 83\% of individual models for cases and 91\% for deaths, with median ensemble aggregation proving more robust than mean aggregation \citep{sherratt2023predictive}. This 
cross-national consistency validates the generalizability of 
ensemble forecasting strategies.
We also observed trade-offs between accuracy and uncertainty quantification. EM3 UW delivered accurate forecasts and maintained the best 95\% PI coverage, highlighting its ability to provide sharp, reliable predictions. These results align with earlier findings from \citep{gneiting2007, funk2019}, which emphasized the importance of well-calibrated forecasts that balance precision with uncertainty.

Model performance varied across regions and forecast horizons. For example, EM3 UW dominated in the Midwest, especially at 3- to 4-week forecast horizons, while GAM and ARIMA excelled at 1- to 2-week-ahead forecasts. In the South and Northeast regions, ARIMA consistently performed well in the short term (1- to 2-week), but ensemble models like EM3 UW and EM2 UW showed more substantial 3- to 4-week reliability. In the West, EM2 UW and EM3 UW performed very well at 3- to 4-week forecasts. These results emphasize the need for context-specific modeling strategies, but the observed heterogeneity should be interpreted cautiously because regional differences in forecast performance may reflect both model behavior and differences in the wastewater surveillance signal itself. Prior wastewater studies have shown that signal interpretation can vary with catchment characteristics, sampling strategy, environmental conditions, and the relationship between wastewater measurements and epidemiologic indicators. \citep{bertels2022factors,zhan2023correlative}

\subsection{Regional Heterogeneity in Forecast Performance}

Our results reveal substantial regional variation in forecast accuracy, with the Northeast consistently exhibiting higher prediction errors across models and horizons. Understanding why certain regions present greater forecasting challenges offers insights into the relationship between epidemic dynamics and predictability.

Several intersecting factors likely contribute to the Northeast's reduced forecast accuracy. First, the region encompasses both dense urban centers (New York City, Boston, Philadelphia) and lower-density suburban areas within the same wastewater catchment aggregations. Epidemic dynamics in high-density settings exhibit more volatile week-to-week fluctuations due to rapid transmission when susceptible populations are depleted and greater sensitivity to behavioral changes. Second, Northeastern cities often operate older combined sewer systems in which stormwater mixes with sewage. During precipitation events, dilution effects can substantially reduce measured WVAL without corresponding reductions in actual infection levels, introducing measurement noise independent of true epidemic dynamics. Third, pronounced seasonality with harsh winters drives indoor congregation and affects both viral shedding rates and RNA degradation in wastewater. This environmental variability compounds forecasting challenges. Finally, substantial non-resident populations (daily commuters, seasonal tourists) contribute to wastewater but may not reflect local residential infection patterns, creating a spatial mismatch between catchments and epidemic dynamics.

An additional consideration is that forecast performance may also have been influenced by the dominant circulating SARS-CoV-2 variant during a given period and region. Variant turnover can alter transmission dynamics, fecal shedding patterns, disease severity, and the relationship between wastewater viral activity and downstream epidemiologic indicators. Wastewater studies have shown that the association between wastewater trends and reported cases or hospitalizations can differ across dominant variant waves, with both correlation strength and lead time varying over time \citep{zhan2023correlative,hill2023wastewater}. Because our forecasting framework did not explicitly incorporate sequencing-informed variant prevalence or variant-specific covariates, these effects may have contributed to the temporal and regional differences in model performance observed here. Future work could integrate genomic surveillance information from public sources of lineage prevalence to assess whether variant-aware models improve forecast performance.

Another important limitation is that we did not explicitly adjust model comparisons for differences in surveillance-network coverage across regions or time. NWSS expanded rapidly during the pandemic, and the number, distribution, and consistency of contributing sampling sites can differ across regions and periods \citep{adams2024national}. As a result, some between-region differences in forecast performance may reflect not only underlying epidemic dynamics or model structure, but also differences in the amount of information contributing to a regional WVAL signal. Future work could evaluate forecast performance jointly with surveillance coverage metrics such as the number of contributing facilities, catchment population represented, or temporal completeness of reporting.

These findings suggest region-specific modeling adaptations. In high-density regions like the Northeast, models that account for rapid behavioral responses (GAM with adaptive smoothing) or capture multiple epidemic phases (\textit{n}-sub-epidemic framework) may outperform simple autoregressive approaches. In contrast, the South and Midwest's more stable dynamics allow simpler models (ARIMA, SLR) to achieve competitive performance with lower computational cost. The West's geographic diversity (urban coast versus rural interior) suggests ensemble approaches combining models with different structural assumptions may provide robustness against this heterogeneity.

The regional performance differences have practical implications for public health surveillance. Decision-makers in the Northeast should anticipate wider forecast uncertainty bounds, and decision protocols should account for this irreducible uncertainty. Regions with volatile dynamics may benefit from increased sampling frequency or more sophisticated protocols (24-hour composite samples, flow-weighted sampling) to reduce measurement noise. Rather than deploying a single national model, our results support a portfolio approach where regional models are tailored to local epidemic characteristics. The consistent superiority of certain models in specific regions suggests local agencies could adopt region-optimized approaches.

\subsection{Reconciling Ensemble and Region-Specific Forecasting}

Our study advocates both ensemble forecasting and region-specific model selection, approaches that might appear contradictory but operate at different levels of the forecasting problem. Ensemble methods address \textit{temporal uncertainty}: at any given forecast time, we cannot know which model will perform best in the upcoming weeks. ARIMA excels during stable periods with predictable autocorrelation; \textit{n}-sub-epidemic models succeed during growth and decline phases when sub-epidemic structure is evident; GAM adapts well to changing smoothness requirements. By combining models, ensembles achieve robustness across varying conditions. Our results demonstrate this: The unweighted ensemble EM3 UW frequently outperforms individual models in most regions, particularly at the 3- and 4-week horizons, by hedging against temporal uncertainty.

Regional adaptation addresses \textit{spatial heterogeneity}: different regions exhibit systematically different epidemic characteristics. The Northeast's volatile dynamics favor models that adapt quickly (GAM), while the South's more stable patterns allow simpler approaches (ARIMA) to perform competitively. However, this spatial heterogeneity may also reflect differences in wastewater surveillance infrastructure, catchment composition, and temporal coverage of contributing sites, which were not explicitly modeled in the present study. Region-specific modeling means selecting which models to include in regional ensembles (perhaps excluding Prophet if it consistently underperforms), adjusting model hyperparameters for regional characteristics (e.g., shorter smoothing windows for volatile regions), and interpreting forecast uncertainty differently across regions based on regional predictability.

We recommend a hierarchical strategy for public health deployment. At the national level, maintain an ensemble of diverse models (ARIMA, GAM, \textit{n}-sub-epidemic, Prophet) to ensure robustness across all regions and time periods. At the regional level, optimize which models to include in regional ensembles, adjust ensemble weighting schemes (equal weights versus performance-based), and tune model-specific hyperparameters. At the temporal monitoring level, continuously assess which models currently perform best and adjust ensemble compositions accordingly. This hierarchical approach leverages both strengths: ensemble robustness protects against model failures, while regional adaptation improves baseline performance.

The temporal performance variation we observe in Figure~\ref{fig:8} reveals an important pattern: model performance deteriorates significantly during periods of rapid change (surges) in viral activity, suggesting potential for adaptive weighting strategies. Recent work demonstrated that adaptive weighted ensembles with exponential time decay (assigning the highest weight to the most recent performance) achieved 52\% error reduction compared to static weighting 
approaches in contexts with irregular epidemic patterns \citep{tsang2024adaptive}. Such approaches could address the performance degradation during surge periods that is evident in our analysis.

Our study period spans a critical surveillance transition:  clinical COVID-19 case reporting was largely discontinued in March 2024, with hospitalization reporting ending in May 2024 \citep{schenk2024sars}. This shift elevates wastewater surveillance 
from complementary to essential status. Our findings validate WBE-based forecasting across this transition, demonstrating stable performance despite changes in clinical reporting infrastructure.

Our findings also suggest a degree of model stability over time, with the top-performing models maintaining consistent ranks across multiple forecasting periods, suggesting that the ensemble models do not exhibit drastic fluctuations in performance.  This consistency is critical for real-time applications, as it suggests these models can remain reliable without frequent recalibration. This is an important consideration for public health use \citep{shaman2013, brooks2018}.

Wastewater surveillance remains a promising tool for early detection of COVID-19 and other infectious diseases. The ability of ensemble models to extract meaningful trends from this data source enhances its value for public health decision-making \citep{peccia2020measurement, wu2021wastewater}. Short-term forecasts (1- to 2-week) can benefit from flexible statistical models like ARIMA and GAM, while long-term (3- to 4-week) projections are best captured by the \textit{n}-sub-epidemic ensemble framework. This strategic allocation of models can support more responsive and informed intervention planning.

Forecast performance in this study was assessed by comparing each model's 1- to 4-week-ahead predictions against subsequently observed WVAL values that were not used in model fitting. Thus, the reported errors and uncertainty metrics reflect out-of-sample predictive performance within the retrospective rolling-origin design, rather than descriptive agreement with data used to estimate the models.

An important practical consideration is the computational cost of different modeling approaches. While the \textbf{n}-sub-epidemic framework requires more computational resources than simple statistical models, the improved forecast accuracy and uncertainty quantification justify this additional cost, especially for public health applications where reliable predictions are critical. Future work could explore more computationally efficient implementations of ensemble methods without sacrificing performance.

Nonetheless, several limitations must be acknowledged. First, this was a retrospective forecasting evaluation, meaning that model predictions were assessed against subsequently observed historical data rather than in a live operational setting. Although this design is useful for comparing predictive performance, it does not fully capture real-time constraints such as reporting delays, data revisions, or abrupt surveillance changes. Second, our analysis was conducted on the published weekly WVAL series and did not explicitly incorporate site-level metadata, surveillance-network coverage, or sequencing-informed variant prevalence. Consequently, some regional and temporal differences in model performance may reflect differences in the number and consistency of contributing sites, variant-dependent shedding and transmission dynamics, or changes in the relationship between wastewater signals and downstream epidemiologic indicators \citep{zhan2023correlative,adams2024national}. Third, wastewater measurements may vary across jurisdictions because of differences in wastewater treatment plant characteristics, catchment composition, sampling frequency, environmental conditions, laboratory workflows, and normalization practices \citep{bertels2022factors}. These factors may influence the comparability of the observed signal across locations and over time. Finally, our modeling framework focused on phenomenological and statistical approaches and did not include mechanistic SEIR-type models or hybrid models that explicitly incorporate transmission mechanisms, interventions, or variant information. Hybrid approaches combining mechanistic structure with ensemble phenomenological forecasting could improve performance during periods of rapidly changing transmission rates.

Our study also did not explicitly stratify model performance by epidemic phase (rising, peak/plateau, declining). Our rolling 10-week calibration framework was designed to mimic operational forecasting use and naturally spans multiple phases, but it does not allow formal phase-specific comparisons without additional assumptions and loss of sample size. Future work could incorporate explicit phase definitions to evaluate performance across distinct stages of epidemic dynamics.

Additionally, while unweighted ensembles performed well, further research is needed to refine ensemble strategies, such as developing adaptive or data-driven weighting schemes. Exploring hybrid approaches that integrate statistical, machine learning, and mechanistic elements could also improve adaptability and accuracy across forecast horizons. Finally, limitations in the wastewater dataset, such as uneven geographic coverage, reporting inconsistencies, and noise in signal detection, may have impacted model performance in ways not fully captured in this study.

\section{Conclusion}
Overall, our findings reinforce the benefits of ensemble forecasting approaches for WVAL, particularly unweighted ensembles, which performed well across multiple metrics. These results emphasize the importance of balancing accuracy and uncertainty quantification when designing epidemic forecasting models. Future research should explore alternative weighting schemes, investigate the impact of model diversity within ensembles, and extend this analysis to additional regions and time periods. Additionally, integrating domain knowledge with machine learning techniques could further enhance the interpretability and robustness of epidemic forecasts \citep{viboud2018rapidd}.

Our study provides actionable insights for public health practitioners: use ARIMA or GAM for immediate (1- to 2-week) operational planning, and rely on \textit{n}-sub-epidemic ensemble models for medium-term (3-4 week) strategic planning. This tiered approach can optimize resource allocation and intervention timing based on forecast reliability at different horizons.

These insights contribute to the ongoing development of effective and reliable forecasting tools for epidemic response and public health decision-making. Future work should focus on optimizing ensemble model structures to maximize predictive performance while ensuring adaptability to different epidemiological contexts. 

\clearpage

\textbf{Author Contributions}\\
F.A. and G.C. conceived the study. F.A., H.K., A.B., and G.C. contributed to the conceptualization and methodology of the study, participated in validation of results, curated the data, wrote the original manuscript
draft, and prepared the visualizations (figures and tables). All contributed to the writing and revising of earlier versions of the manuscript.

\textbf{Funding}\\
 G.C. was supported by NSF Awards 2125246 and OISE-2412914. F.A. is supported through a 2CI Fellowship from Georgia State University.


\clearpage
\bibliographystyle{plainnat}

\bibliography{myreference}

\clearpage
\setcounter{figure}{0}
\renewcommand{\thefigure}{S\arabic{figure}}

\setcounter{table}{0}
\renewcommand{\thetable}{S\arabic{table}}

\section{Supplementary}

\begin{figure}[H]
    \centering
    \includegraphics[width=1\linewidth]{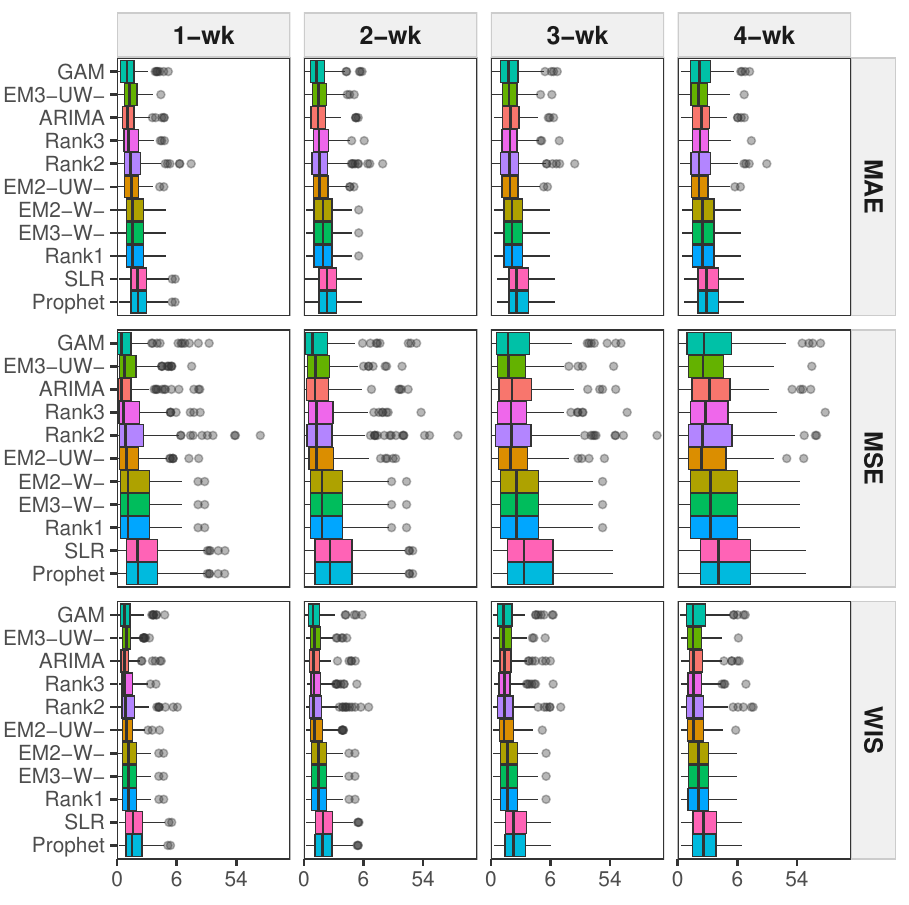}
    \caption{National-level distributions of forecast performance across models and horizons. Boxplots show the distributions of MAE, MSE, and WIS (rows; plotted on a log scale) for 1–4 week-ahead forecasts (columns) over all national forecast dates. For each model, the vertical line inside the box marks the median, and the box width spans the interquartile range. Whiskers extend to 1.5×IQR, and gray points indicate outliers.}
    \label{S1}
\end{figure}

\begin{figure}[H]
    \centering
    \includegraphics[width=.75\textwidth]{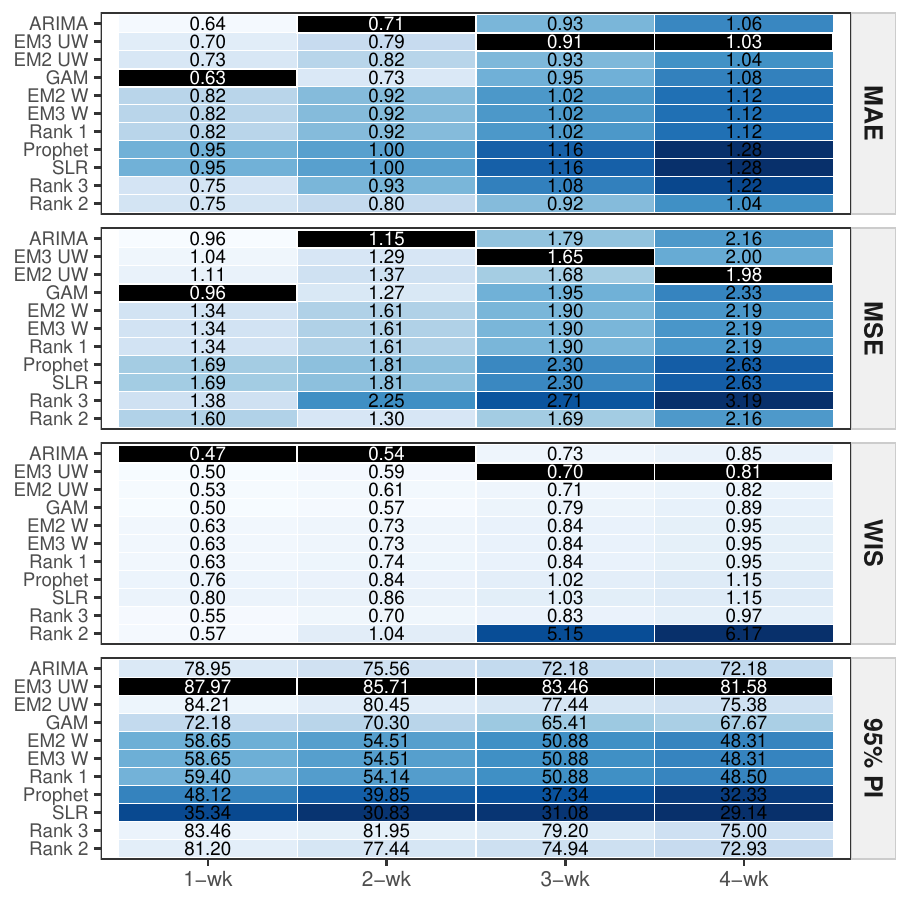}
    \caption{\footnotesize{Log-transformed averages of MAE, MSE, and WIS, and the (untransformed) average 95\% PI coverage across all models for 1--4 week horizons in the Midwest region, spanning the period from March 5, 2022, to September 14, 2024 (133 forecasts). 
Lighter shades indicate smaller (better) values for MAE, MSE, and WIS, whereas darker shades indicate larger errors. 
For coverage, values closer to 95\% indicate better performance. 
Black cells highlight the best-performing model(s) for each forecast horizon and metric.
}}
    \label{fig:5}
\end{figure}

\begin{figure}[H]
    \centering
    \includegraphics[width=1\linewidth]{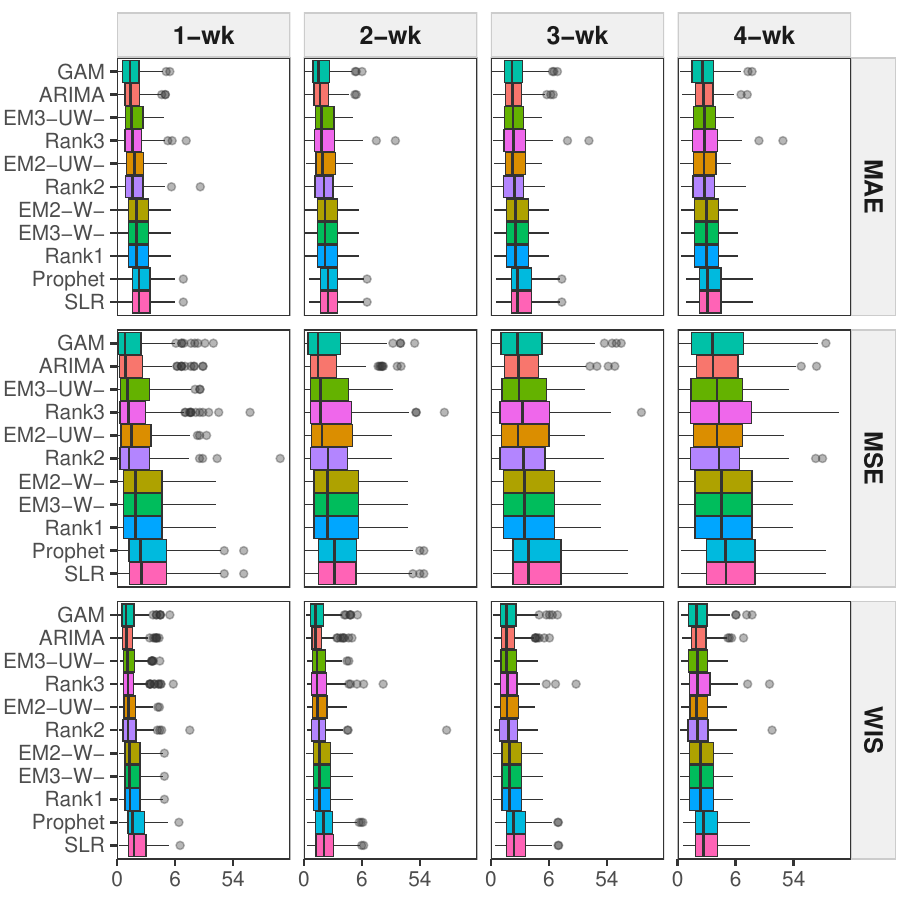}
    \caption{Midwest region distributions of forecast performance across models and horizons. Boxplots show the distributions of MAE, MSE, and WIS (rows; plotted on a log scale) for 1–4 week-ahead forecasts (columns) over all Midwest region forecast dates. For each model, the vertical line inside the box marks the median, and the box width spans the interquartile range. Whiskers extend to 1.5×IQR, and gray points indicate outliers.}
    \label{S2}
\end{figure}

\begin{figure}[H]
    \centering
    \includegraphics[width=.75\textwidth]{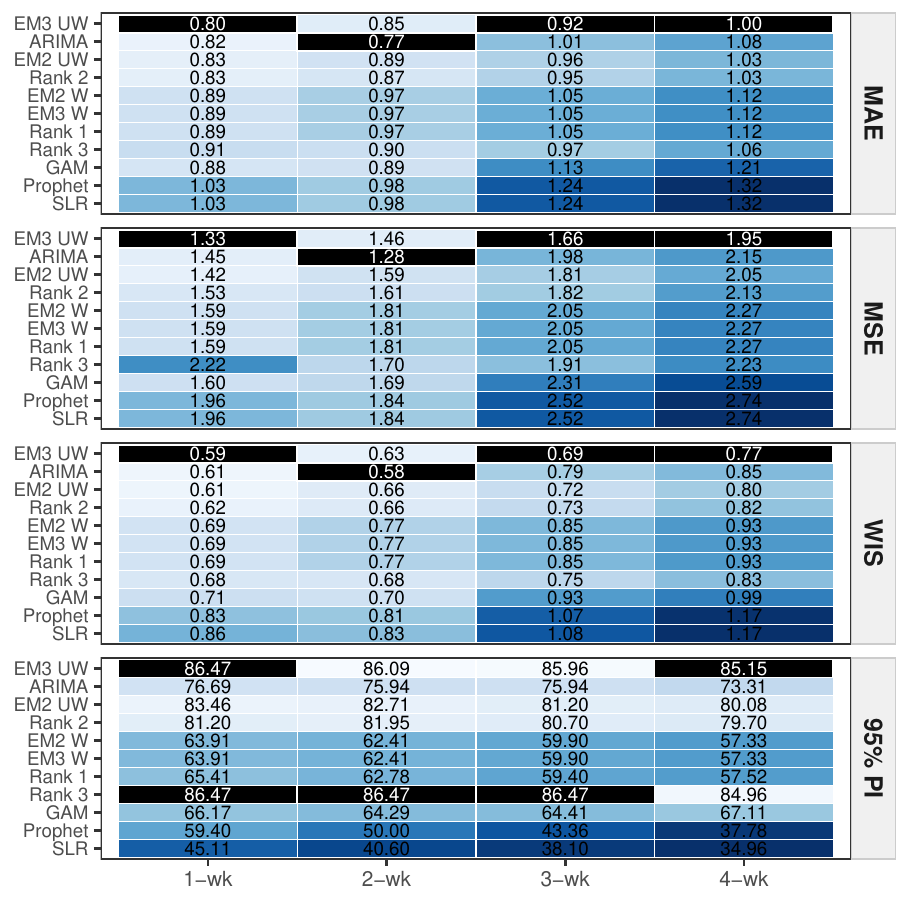}
    \caption{\footnotesize{Log-transformed averages of MAE, MSE, and WIS, and the (untransformed) average 95\% PI coverage across all models for 1--4 week horizons in the Northeast region, spanning the period from March 5, 2022, to September 14, 2024 (133 forecasts). 
Lighter shades indicate smaller (better) values for MAE, MSE, and WIS, whereas darker shades indicate larger errors. 
For coverage, values closer to 95\% indicate better performance. 
Black cells highlight the best-performing model(s) for each forecast horizon and metric.
}}
    \label{fig:6}
\end{figure}

\begin{figure}[H]
    \centering
    \includegraphics[width=1\linewidth]{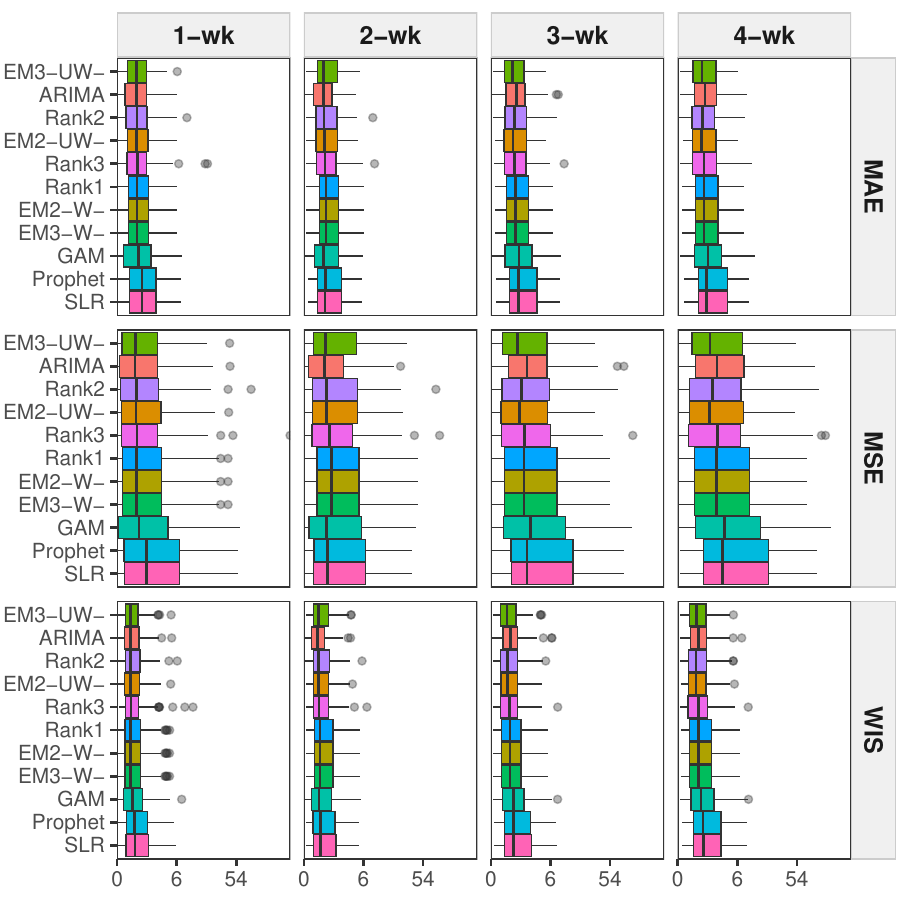}
    \caption{Northeast region distributions of forecast performance across models and horizons. Boxplots show the distributions of MAE, MSE, and WIS (rows; plotted on a log scale) for 1–4 week-ahead forecasts (columns) over all Northeast region forecast dates. For each model, the vertical line inside the box marks the median, and the box width spans the interquartile range. Whiskers extend to 1.5×IQR, and gray points indicate outliers. }
    \label{S3}
\end{figure}

\begin{figure}[H]
    \centering
    \includegraphics[width=.75\textwidth]{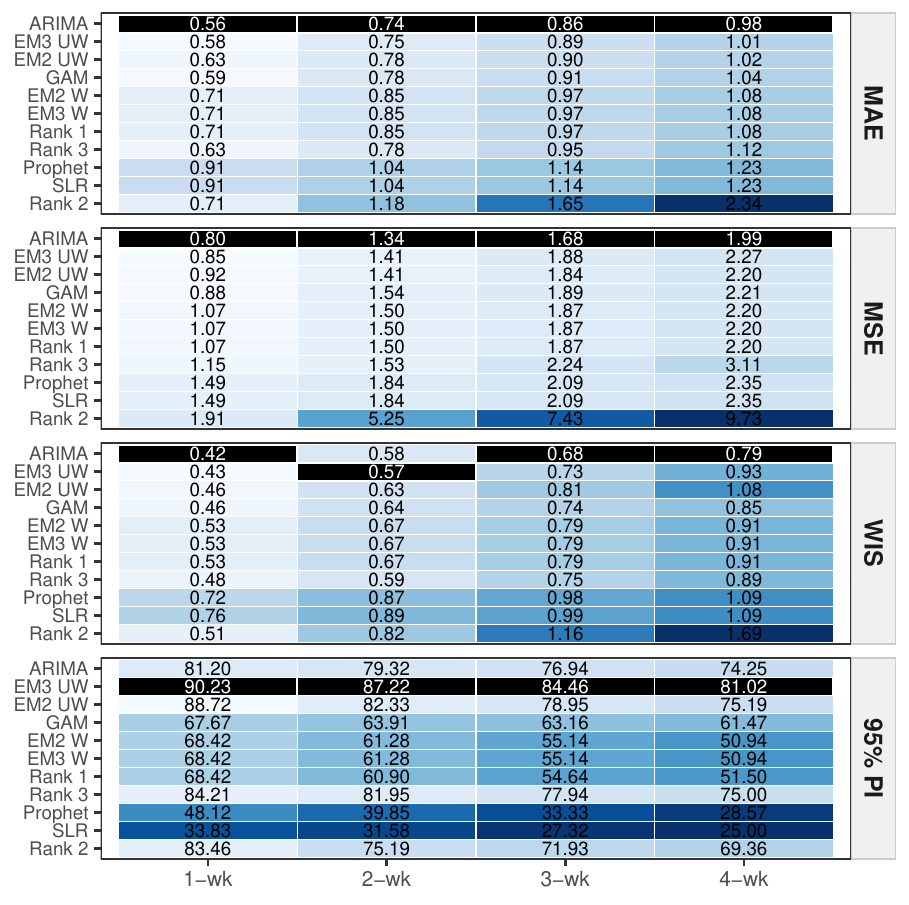}
    \caption{\footnotesize{Log-transformed averages of MAE, MSE, and WIS, and the (untransformed) average 95\% PI coverage across all models for 1--4 week horizons in the South region, spanning the period from March 5, 2022, to September 14, 2024 (133 forecasts). 
Lighter shades indicate smaller (better) values for MAE, MSE, and WIS, whereas darker shades indicate larger errors. For coverage, values closer to 95\% indicate better performance.
Black cells highlight the best-performing model(s) for each forecast horizon and metric.
}}
    \label{fig:7}
\end{figure}

\begin{figure}[H]
    \centering
    \includegraphics[width=1\linewidth]{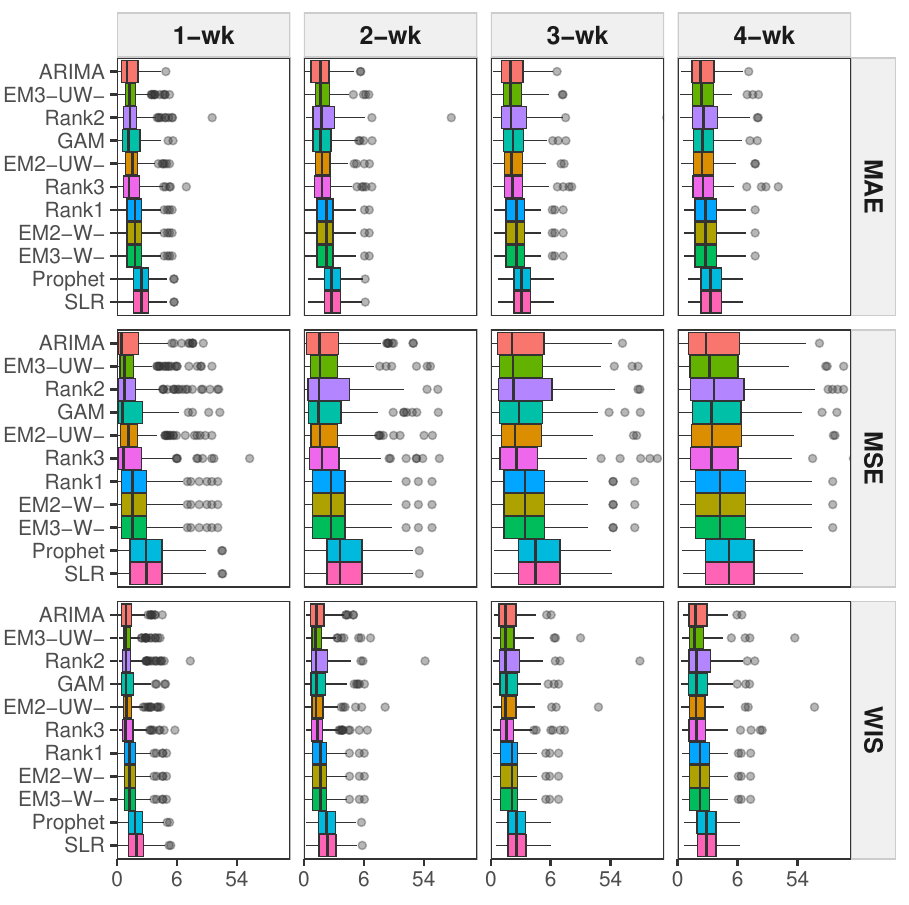}
    \caption{South region distributions of forecast performance across models and horizons. Boxplots show the distributions of MAE, MSE, and WIS (rows; plotted on a log scale) for 1–4 week-ahead forecasts (columns) over all South region forecast dates. For each model, the vertical line inside the box marks the median, and the box width spans the interquartile range. Whiskers extend to 1.5×IQR, and gray points indicate outliers.}
    \label{S4}
\end{figure}

\begin{figure}[H]
    \centering
    \includegraphics[width=.75\textwidth]{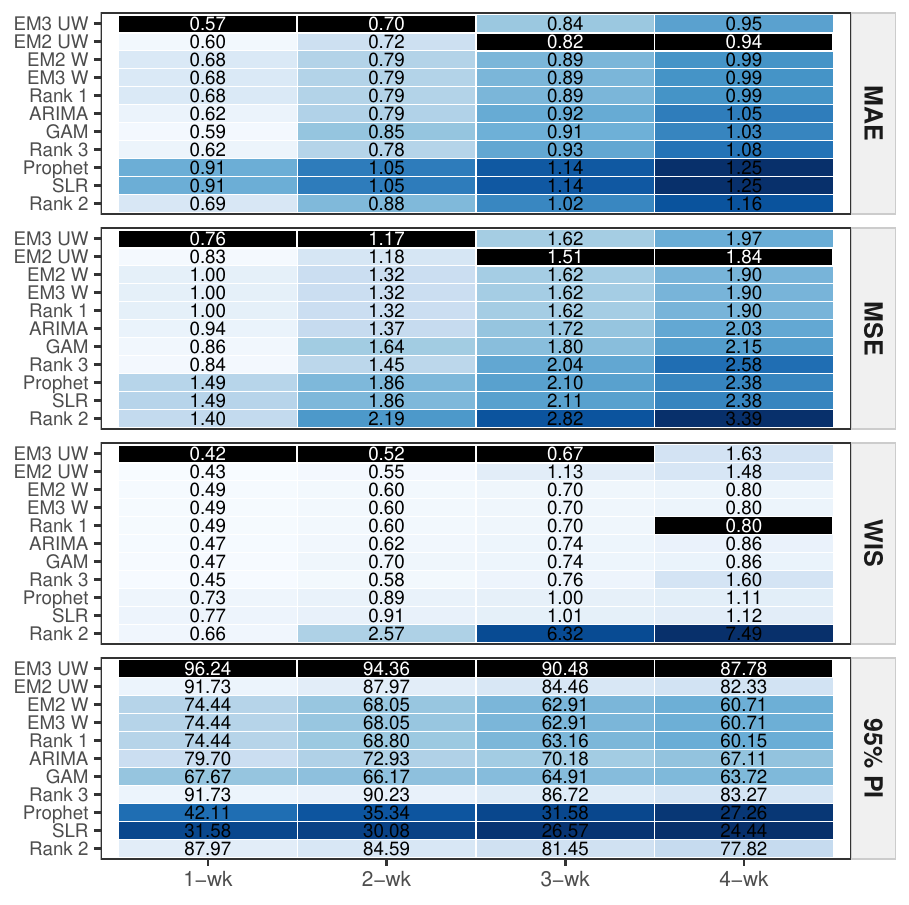}
    \caption{\footnotesize{Log-transformed averages of MAE, MSE, and WIS, and the (untransformed) average 95\% PI coverage across all models for 1--4 week horizons in the West region, spanning the period from March 5, 2022, to September 14, 2024 (133 forecasts). 
Lighter shades indicate smaller (better) values for MAE, MSE, and WIS, whereas darker shades indicate larger errors. For coverage, values closer to 95\% indicate better performance. Black cells highlight the best-performing model(s) for each forecast horizon and metric.
}}
    \label{fig:new}
\end{figure}
\begin{figure}[H]
    \centering
    \includegraphics[width=1\linewidth]{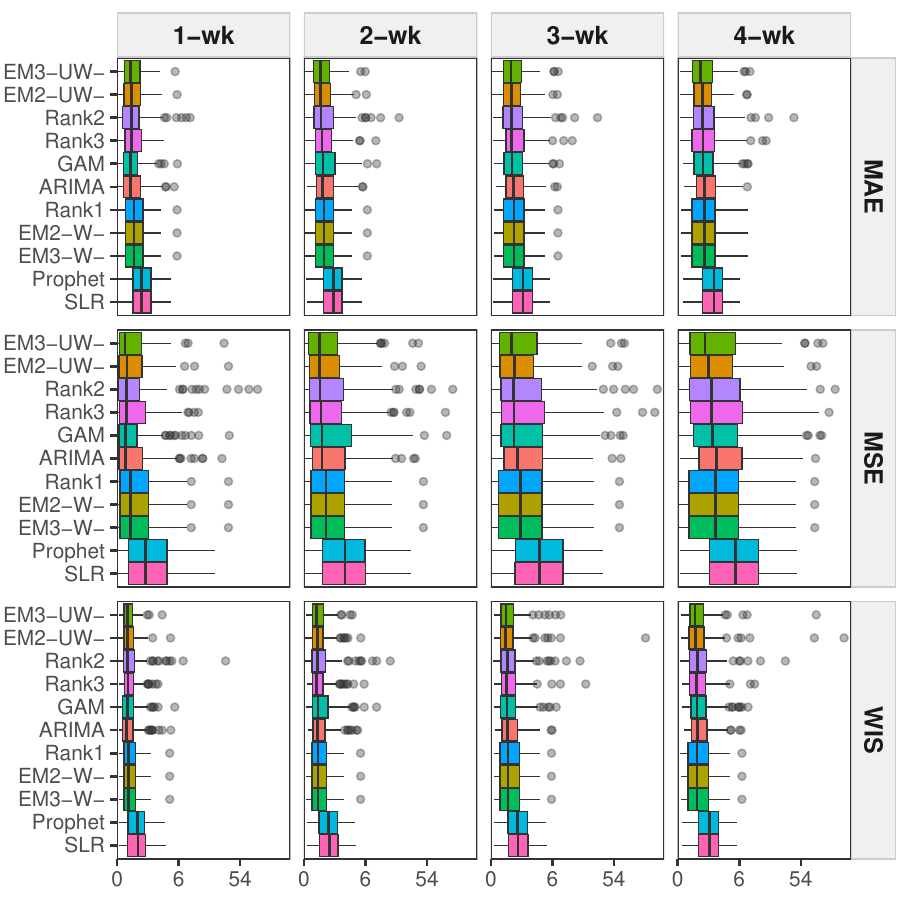}
    \caption{West region distributions of forecast performance across models and horizons. Boxplots show the distributions of MAE, MSE, and WIS (rows; plotted on a log scale) for 1–4 week-ahead forecasts (columns) over all West region forecast dates. For each model, the vertical line inside the box marks the median, and the box width spans the interquartile range. Whiskers extend to 1.5×IQR, and gray points indicate outliers.}
    \label{S5}
\end{figure}

\begin{figure}[H]
    \centering
    \includegraphics[width=1\linewidth]{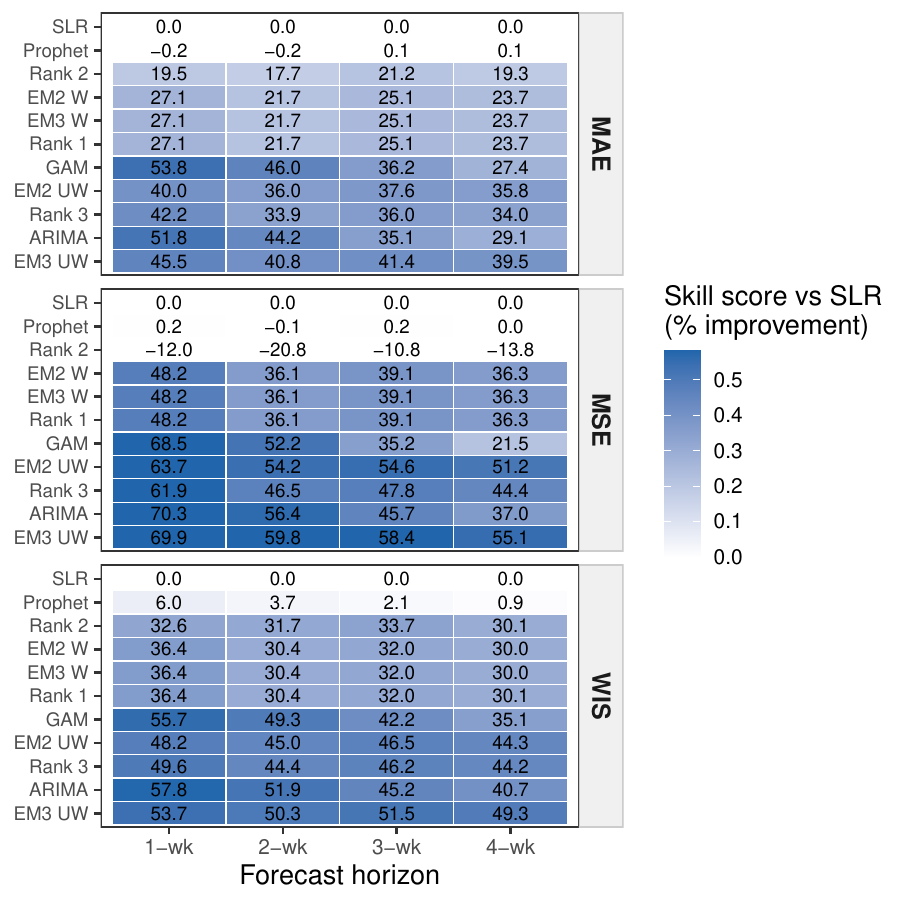}
    \caption{Skill scores relative to the SLR baseline at the national level across models, horizons, and metrics. Heatmaps display percent skill scores ($100 \times [1 - \bar{m} / \bar{m}_{\text{SLR}}]$), where $\bar{m}$ and $\bar{m}_{\text{SLR}}$ denote the average MAE, MSE, or WIS across all forecast dates for a given model and for the SLR baseline, respectively. Rows correspond to MAE, MSE, and WIS, and columns to 1--4 week-ahead horizons. Positive values (blue) indicate lower error than SLR (better performance), while values near zero (white) indicate performance similar to the baseline; SLR is therefore zero by construction. The fill scale was bounded using the 5th and 95th percentiles of the skill-score distribution, with values outside this range clipped to the endpoint colors, while cell annotations report the original percent skill values.}
    \label{S6}
\end{figure}

\begin{figure}[H]
    \centering
    \includegraphics[width=1\linewidth]{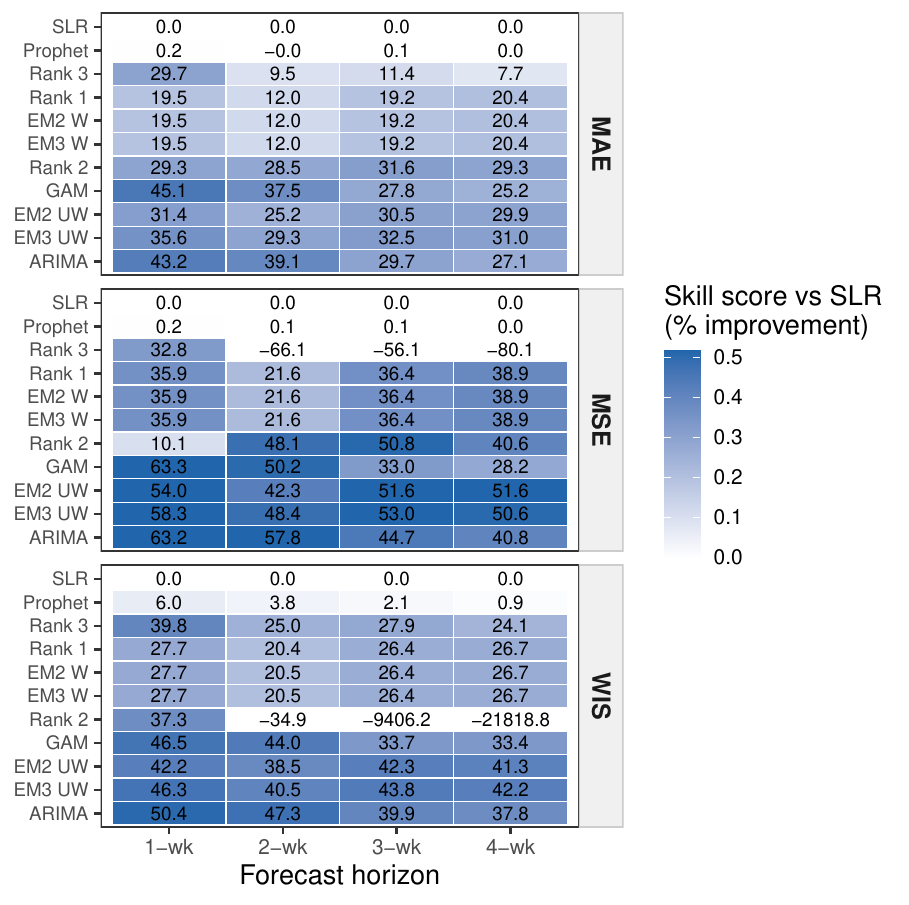}
    \caption{Midwest region skill scores relative to the SLR baseline across models, horizons, and metrics. Heatmaps display percent skill scores ($100 \times [1 - \bar{m} / \bar{m}_{\text{SLR}}]$), where $\bar{m}$ and $\bar{m}_{\text{SLR}}$ denote the average MAE, MSE, or WIS across all forecast dates for a given model and for the SLR baseline, respectively. Rows correspond to MAE, MSE, and WIS, and columns to 1--4 week-ahead horizons. Positive values (blue) indicate lower error than SLR (better performance), while values near zero (white) indicate performance similar to the baseline; SLR is therefore zero by construction. The fill scale was bounded using the 5th and 95th percentiles of the skill-score distribution, with values outside this range clipped to the endpoint colors, while cell annotations report the original percent skill values.}

    \label{S7}
\end{figure}

\begin{figure}[H]
    \centering
    \includegraphics[width=1\linewidth]{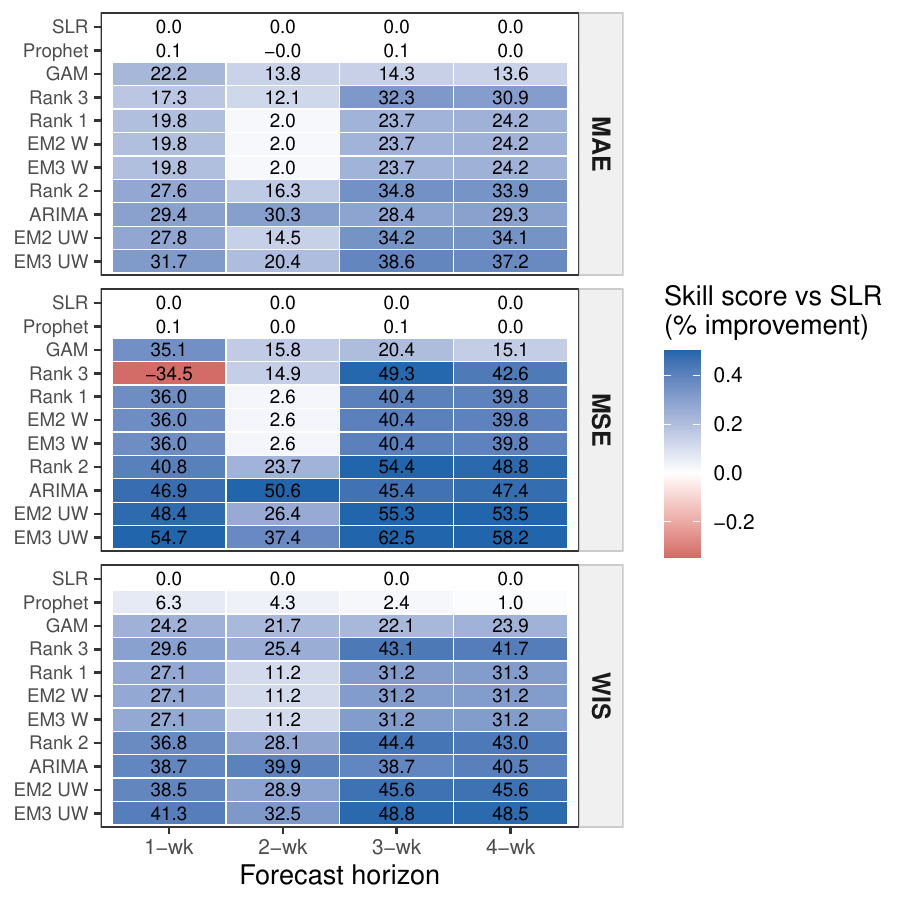}
    \caption{Northeast region skill scores relative to the SLR baseline across models, horizons, and metrics. Heatmaps display percent skill scores ($100 \times [1 - \bar{m} / \bar{m}_{\text{SLR}}]$), where $\bar{m}$ and $\bar{m}_{\text{SLR}}$ denote the average MAE, MSE, or WIS across all forecast dates for a given model and for the SLR baseline, respectively. Rows correspond to MAE, MSE, and WIS, and columns to 1--4 week-ahead horizons. Positive values (blue) indicate lower error than SLR (better performance), whereas negative values (red) indicate higher error than SLR (worse performance); by definition, SLR has a skill score of zero. To reduce the influence of extreme negative outliers on the color mapping, the fill scale was bounded using the 5th and 95th percentiles of the skill-score distribution: values below the 5th percentile and above the 95th percentile were clipped to the endpoint colors, while cell annotations report the original percent skill values.}

    \label{S8}
\end{figure}

\begin{figure}[H]
    \centering
    \includegraphics[width=1\linewidth]{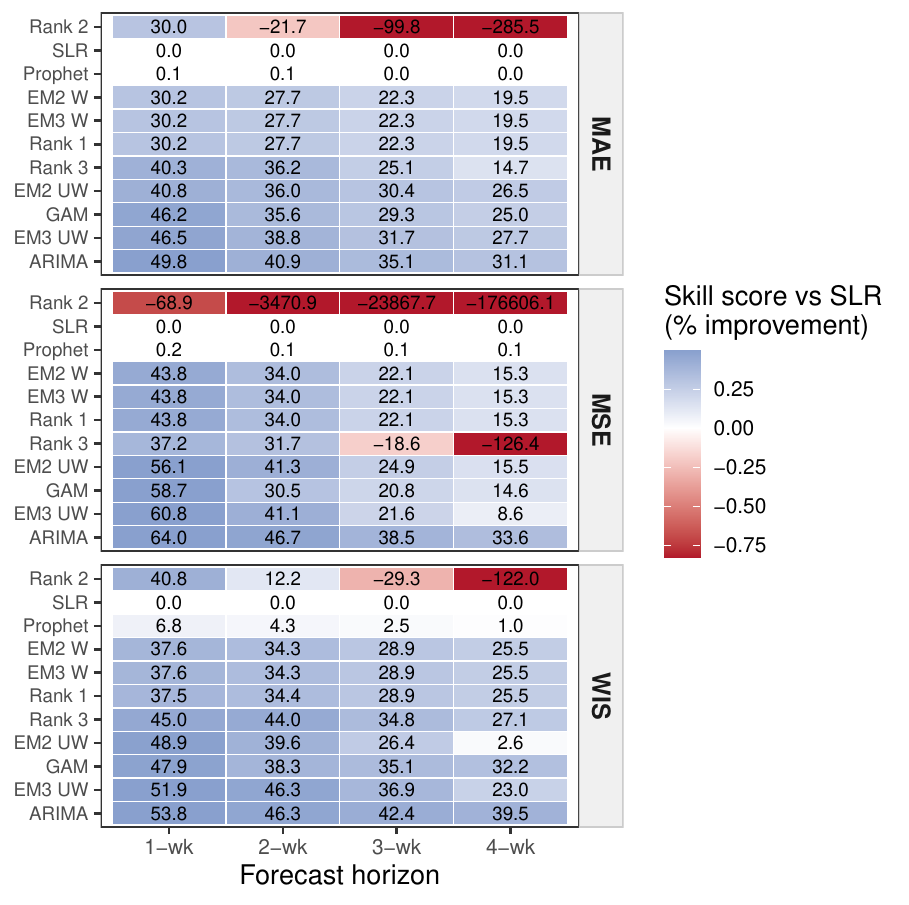}
    \caption{South region skill scores relative to the SLR baseline across models, horizons, and metrics. Heatmaps display percent skill scores ($100 \times [1 - \bar{m} / \bar{m}_{\text{SLR}}]$), where $\bar{m}$ and $\bar{m}_{\text{SLR}}$ denote the average MAE, MSE, or WIS across all forecast dates for a given model and for the SLR baseline, respectively. Rows correspond to MAE, MSE, and WIS, and columns to 1--4 week-ahead horizons. Positive values (blue) indicate lower error than SLR (better performance), whereas negative values (red) indicate higher error than SLR (worse performance); by definition, SLR has a skill score of zero. To reduce the influence of extreme negative outliers on the color mapping, the fill scale was bounded using the 5th and 95th percentiles of the skill-score distribution: values below the 5th percentile and above the 95th percentile were clipped to the endpoint colors, while cell annotations report the original percent skill values.}
    \label{S9}
\end{figure}

\begin{figure}[H]
    \centering
    \includegraphics[width=1\linewidth]{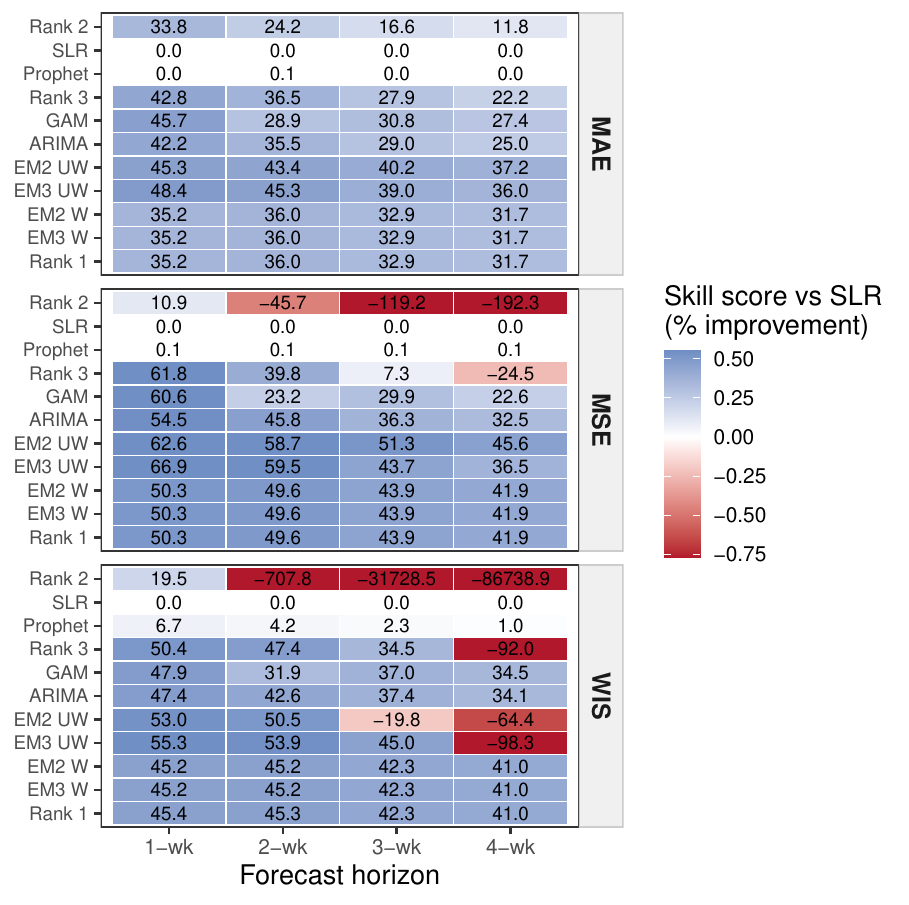}
    \caption{West region skill scores relative to the SLR baseline across models, horizons, and metrics. Heatmaps display percent skill scores ($100 \times [1 - \bar{m} / \bar{m}_{\text{SLR}}]$), where $\bar{m}$ and $\bar{m}_{\text{SLR}}$ denote the average MAE, MSE, or WIS across all forecast dates for a given model and for the SLR baseline, respectively. Rows correspond to MAE, MSE, and WIS, and columns to 1--4 week-ahead horizons. Positive values (blue) indicate lower error than SLR (better performance), whereas negative values (red) indicate higher error than SLR (worse performance); by definition, SLR has a skill score of zero. To reduce the influence of extreme negative outliers on the color mapping, the fill scale was bounded using the 5th and 95th percentiles of the skill-score distribution: values below the 5th percentile and above the 95th percentile were clipped to the endpoint colors, while cell annotations report the original percent skill values.}
    \label{S10}
\end{figure}

\end{document}